\documentclass[12pt,draft, onecolumn]{./IEEEtran}
\usepackage{xspace,amsmath,amssymb,epsfig,syntonly}
\renewcommand{\baselinestretch}{1.65}
\psfull  

\newtheorem{algorithm}{\hspace{-11pt}\bf Algorithm}
\newtheorem{lemma}{\hspace{-11pt}\bf Lemma}

\newtheorem{proposition}{\hspace{-11pt}\bf Proposition}
\newtheorem{theorem}{\hspace{-11pt}\bf Theorem}
\newtheorem{corollary}{\hspace{-11pt}\bf Corollary}

\newtheorem{definition}{\hspace{-11pt}\bf Definition}

\begin{document}
\title{\Large\sf\hspace{1\linewidth} Energy-Efficient Resource Allocation in Time Division
Multiple-Access over Fading Channels$^*$ \vspace{.5in}}
\author{
   {\normalsize\it \large Xin Wang}
   {\normalsize \it \large and Georgios B. Giannakis$^\dagger$
                   {\sf (contact author)}}\\[5cm]
   \thanks{\protect\rule{0pt}{2cm}$^*$
           Work in this paper was supported by the ARO Grant No. W911NF-05-1-0283
           and was prepared through collaborative
           participation in the Communications and Networks Consortium
           sponsored by the U. S. Army Research Laboratory under the
           Collaborative Technology Alliance Program, Cooperative Agreement
           DAAD19-01-2-0011. The U. S. Government is authorized to
           reproduce and distribute reprints for Government purposes
           notwithstanding any copyright notation thereon.
   }
   \thanks{\protect\rule{0pt}{1.5em}$\dagger$ The authors are
           with the Dept. of Electrical and Computer Engineering,
           University of Minnesota, 200 Union Street SE, Minneapolis,
           MN 55455. Tel/fax: (612)626-7781/625-4583;
           Email:~{\tt \{xinwest,georgios\}{\rm\char64}ece.umn.edu}.
    }
   {\small
      \begin{center}
        \[
           \begin{array}{rl}
               \text{\bf Submission date:} & \text{\today}\\
               \text{\bf Suggested Editorial Areas:} & \text{Multi-Access Fading Channels}\\
                                          & \text{Optimal Resource Allocation}\\
           \end{array}
         \]
      \end{center}
    }
}

\markboth{IEEE Transactions on Information Theory (submitted)}
         {}

\renewcommand{\thepage}{}
\maketitle \pagestyle{plain} \pagenumbering{arabic}

\newpage
\setcounter{page}{1}
\begin{abstract}
We investigate energy-efficiency issues and resource allocation
policies for time division multi-access (TDMA) over fading
channels in the power-limited regime. Supposing that the channels
are frequency-flat block-fading and transmitters have full or
quantized channel state information (CSI), we first minimize power
under a weighted sum-rate constraint and show that the optimal
rate and time allocation policies can be obtained by water-filling
over realizations of convex envelopes of the minima for
cost-reward functions. We then address a related minimization
under individual rate constraints and derive the optimal
allocation policies via greedy water-filling. Using water-filling
across frequencies and fading states, we also extend our results
to frequency-selective channels. Our approaches not only provide
fundamental power limits when each user can support an infinite
number of capacity-achieving codebooks, but also yield guidelines
for practical designs where users can only support a finite number
of adaptive modulation and coding (AMC) modes with prescribed
symbol error probabilities, and also for systems where only
discrete-time allocations are allowed.


\end{abstract}


\section{Introduction}\label{S:intro}

With battery operated communicating nodes, energy efficiency has
emerged as a critical issue in both commercial and tactical radios
designed to extend battery lifetime, especially for wireless
networks of sensors equipped with non-rechargeable batteries.
Because the energy required to transmit a certain amount of
information is an increasing and strictly convex function of the
transmission rate~\cite{Berry02}, energy-efficient resource
allocation has attracted growing
attention~\cite{Uysal-Biyikoglu02net}-\cite{Antonio2005}. Among
them, \cite{Uysal-Biyikoglu02net, ElGamal02, Khojastepour04,
Zafer&Modiano2005} dealt with energy-efficient designs based on
packet arrival and delay constraints over additive white Gaussian
noise (AWGN) channels; while \cite{Fu&Modiano2003} and
\cite{Yao&Giannakis2005} considered energy-efficient scheduling
for time division multi-access (TDMA) networks over fading
channels, where the data of each user must be transmitted by a
given deadline. Recently, \cite{Antonio2005} minimized transmit
power of orthogonal frequency-division multiplexing (OFDM) systems
using quantized channel state information (CSI) of the underlying
fading channel.

Resource allocation for fading channels also remains a popular
topic in information theoretic studies. However, optimization has
been typically carried out to maximize rate (achieve capacity)
subject to average power constraints. Assuming that both
transmitters and receivers have available perfect CSI, Tse and
Hanly derived the ergodic capacity~\cite{Tse&Hanly1998} as well as
the delay-limited capacity regions~\cite{Hanly&Tse1998} along with
the optimal power allocation for fading multi-access channels,
while Li and Goldsmith found the ergodic~\cite{Li&Goldsmith2001a}
and outage capacity regions~\cite{Li&Goldsmith2001b} as well as
optimal resource allocation policies for code division (CD), time
division (TD) and frequency division (FD) fading broadcast
channels. As regarding delay-limited capacity (a.k.a.
``zero-outage capacity''), \cite{Caire1999} and
\cite{Li&Jindal&Goldsmith2005} extended the results
of~\cite{Hanly&Tse1998}, to characterize the outage capacity
regions for single-user fading channels and multi-access fading
channels, respectively.

In this paper, we re-consider these information theoretic results
pertaining to {\it rate} efficiency and investigate optimal
resource allocation for fading channels from an {\it energy}
efficiency perspective. Specifically, we seek to minimize
energy/power cost under average rate constraints for TDMA fading
channels, given perfect or quantized CSI, at the transmit- and
receive-ends. As stated in~\cite{Li&Goldsmith2001a}, TD and FD are
equivalent in the sense that they exhibit identical ergodic
capacity regions and corresponding optimal resource allocation
policies. Thus, our results apply also to FDMA fading channels.
Unlike \cite{Uysal-Biyikoglu02net}-\cite{Yao&Giannakis2005}, we do
not impose delay constraints in our energy minimization problems.

We first study the problem of minimizing total power given a
weighted average sum-rate constraint for the block flat-fading
TDMA channel (Section \ref{S:wsum}). This is dual
to~\cite{Li&Goldsmith2001a}, where rate was maximized under a
sum-power constraint. Note that we impose a {\it weighted}
sum-rate constraint for the multi-access channels while
\cite{Li&Goldsmith2001a} consider the sum-power constraint for the
broadcast channel. We then optimize energy-efficiency when each
user can only support a finite number of adaptive modulation and
coding (AMC) modes. The second problem we consider is power
minimization under individual rate constraints, which is the
general case for multi-access channels (Section \ref{S:ind}). Rate
maximization under individual power constraints has been addressed
via superposition coding and successive decoding
in~\cite{Tse&Hanly1998} and \cite{Hanly&Tse1998}. Here we
formulate and solve energy minimization under individual rate
constraints for TDMA when users have infinite-codebooks, or, a
finite number of AMC-modes which requires only {\it quantized} CSI
to be fed back from the receiver to the transmitters. Section
\ref{S:simul} provides some numerical results, followed by the
conclusions of this paper.

\section{Modeling Preliminaries}\label{S:system}

We consider a set of $K$ users linked wirelessly to a single
access point and adopt a discrete-time multi-access Gaussian
channel model as in~\cite{Tse&Hanly1998, Li&Jindal&Goldsmith2005}:
\begin{equation}\label{E:model}
    y(n)=\sum_{k=1}^{K} \sqrt{h_k(n)} x_k(n) + z(n),
\end{equation}
where $x_k(n)$ and $h_k(n)$ are the transmitted signal and the
fading process of the $k$th user, respectively, and $z(n)$ denotes
AWGN with variance $\sigma^2$. Different from \cite{Tse&Hanly1998}
and \cite{Li&Jindal&Goldsmith2005}, we confine ourselves to TDMA
where each user transmits in a dedicated time fraction, not
overlapping with other users; i.e., when $x_k(n) \neq 0$ in
(\ref{E:model}), we have $x_i(n)=0$ for $\forall i\neq k$. We also
assume that the fading processes of all users are jointly
stationary and ergodic, with continuous stationary
distribution.\footnote{As with \cite{Li&Jindal&Goldsmith2005}, our
analysis can be easily extended to discrete distributions.} The
joint fading process is slowly time-varying relative to the
codeword's length, and adheres to a block fading channel model,
which remains constant for a time block $T$, but is allowed to
change in an independent identically distributed (i.i.d.) fashion
from block to block. This is a valid model for ideally interleaved
TDMA or packet-based access where each data frame ``sees'' an
independent channel realization which remains constant within each
frame~\cite[Chapter 2]{FWC}. User transmissions to the access
point are naturally frame-based, where the frame length is chosen
equal to the block length. Having perfect knowledge of the
(possibly quantized) $\{h_k\}_{k=1}^K$, the access point assigns
time fractions to users via a (uplink map) message before an
uplink frame. Then users transmit with the rate adapted to their
CSI at the assigned time fractions. Let
$\mathbf{h}:=[h_1,\ldots,h_K]^T$ denote the joint fading state
over a block. Through feedback from the access point, the $K$
transmitters are assumed to know $\mathbf{h}$ and can vary their
codewords, transmission rates and transmission times per block.

{\emph Notation:} We use boldface lower-case letters to denote
column vectors and inequalities for vectors are defined
element-wise. We let $F(\mathbf{h})$ denote the cumulative
distribution function (cdf) of joint fading states,
$E_\mathbf{h}[\cdot]$ the expectation operator over fading states,
$f^{(k)}(x)$ the $k$th derivative of $f(x)$, $\phi$ the empty set,
$^T$ the transposition operator, $\mathbf{I}_{\{\cdot\}}$ the
indicator function ($\mathbf{I}_{\{x\}}=1$ if $x$ is true and zero
otherwise), and $[x]_+:=\max (x,0)$.

\section{Weighted Sum Average Rate Constraint}\label{S:wsum}

We first consider the problem of minimizing total power given a
weighted sum average rate constraint. Such a constraint may arise
in a wireless sensor network, where the fusion center requires an
aggregate rate $\bar{R}$ to perform a certain task (e.g.,
distributed estimation) using data from different users with
different reward weights. Given a rate allocation policy
$\mathbf{r}(\cdot)$ and a time allocation policy
$\boldsymbol{\tau}(\cdot)$, let $\tau_k(\mathbf{h})$ and
$r_k(\mathbf{h})$ denote the time {\it fraction} allocated to user
$k$ and the corresponding transmission rate during
$\tau_k(\mathbf{h})$. Taking into account that user $k$ does not
transmit over the remaining $1-\tau_k(\mathbf{h})$ fraction of
time, the $k$th user's {\it overall} transmission rate {\it per
block} is $\tau_k(\mathbf{h}) r_k(\mathbf{h})$. If
$\mathbf{w}:=[w_1, \ldots, w_K]^T$ collects the rate reward
weights assigned to the $K$ users, we let ${\cal F}_{\mathbf{w}}$
denote the set of all possible rate and time allocation policies
satisfying the average rate constraint $E_\mathbf{h}[\sum_{k=1}^K
w_k \tau_k(\mathbf{h}) r_k(\mathbf{h})] \geq \bar{R}$ with
$\sum_{k=1}^K \tau_k(\mathbf{h}) =1$, $\forall \mathbf{h}$.
Clearly, using transmit power $p_k(\mathbf{h})$ during
$\tau_k(\mathbf{h})$ fraction of time in any given block, user $k$
can theoretically transmit with arbitrarily small error a number
of bits/sec up to the Shannon capacity $r_k(\mathbf{h}) = B \log
\left(1+\frac{h_k p_k(\mathbf{h})}{\sigma^2}\right)$, where $B$ is
the system bandwidth. Without loss of generality (w.l.o.g.), we
assume henceforth that $B=1$ and $\sigma^2=1$. Again, notice that
with allocated time fraction $\tau_k(\mathbf{h})$, the $k$th
user's {\it overall} transmit power per block is
$P_k(\mathbf{h})=\tau_k(\mathbf{h}) p_k(\mathbf{h})$ since no
power is used for $1-\tau_k(\mathbf{h})$ fraction of time.

With $\bar{P}_k:= E_\mathbf{h} [P_k(\mathbf{h})]$,
$\mathbf{\bar{p}} := [\bar{P}_1,\ldots, \bar{P}_K]^T$ and in
accordance with the definition of the ergodic capacity region, we
define a {\it power region} as follows.

\begin{definition}
{\it The power region for the TDMA fading channel when
transmitters and the receiver have perfect CSI, is given by
\begin{equation}\label{E:power}
    {\cal P}(\bar{R}) = \bigcup_{(\mathbf{r}(\cdot),\boldsymbol{\tau}(\cdot)) \in {\cal
    F}_{\mathbf{w}}} {\cal P}_{TD} (\mathbf{r}(\cdot),\boldsymbol{\tau}(\cdot)),
\end{equation}
where
\begin{equation}\label{E:PTD}
    {\cal P}_{TD} (\mathbf{r}(\cdot),\boldsymbol{\tau}(\cdot)) = \left\{ \mathbf{\bar{p}}:
    \bar{P}_k
    \geq E_\mathbf{h}\left[ \frac{\tau_k(\mathbf{h})}{h_k}
    \left(2^{r_k(\mathbf{h})}-1 \right)\right], \quad \quad 1 \leq k \leq
    K
    \right\}.
\end{equation}}
\end{definition}

If the block length is sufficiently large and the users are
allowed to use different codewords for different fading states, it
is easy to show that every $\mathbf{\bar{p}} \in {\cal
P}(\bar{R})$ is feasible. Moreover, by the time-sharing argument,
we can show that the $K$-dimensional power region ${\cal
P}(\bar{R})$ is convex in $\mathbf{\bar{p}}$ (the proof mimics the
steps in capacity region derivations~\cite{Tse&Hanly1998,
Li&Goldsmith2001a}, and is omitted for brevity).

Supposing that we assign to users different weights
$\boldsymbol{\mu}:=[\mu_1, \ldots, \mu_K]^T \geq \mathbf{0}$, the
energy-efficient resource allocation problem can be formulated as
\begin{equation}\label{E:wsum}
    \min_{\mathbf{\bar{p}}} \boldsymbol{\mu}^T \mathbf{\bar{p}}, \quad \quad
    \mbox{subject to} \quad  \mathbf{\bar{p}} \in {\cal
    P}(\bar{R}).
\end{equation}
Its solution $\mathbf{\bar{p}}$ yields the optimal rate and time
allocation policies, and lies on the boundary surface of ${\cal
P}(\bar{R})$ due to its convexity. By solving (\ref{E:wsum}) for
all $\boldsymbol{\mu} \geq \mathbf{0}$, we can determine all the
boundary points, and thus the entire power region ${\cal
P}(\bar{R})$. When one or more of the entries of
$\boldsymbol{\mu}$ are zero, the solution to (\ref{E:wsum})
corresponds to an extreme point of the boundary surface of ${\cal
P}(\bar{R})$. By letting some of the weights approach 0, we can
get arbitrarily close to these extreme points. One can refer
to~\cite{Tse&Hanly1998, Hanly&Tse1998, Li&Goldsmith2001a,
Li&Goldsmith2001b, Li&Jindal&Goldsmith2005} for the explicit
characterization of the extreme points.

\subsection{Full CSI and Infinite-Codebooks}

When a user can vary its codebook according to each fading state,
the boundary of ${\cal P}(\bar{R})$ is feasible. Therefore, the
problem (\ref{E:wsum}) can be rewritten as
\begin{equation}\label{E:eq5}
    \begin{cases}
        \min_{\mathbf{r}(\cdot),\boldsymbol{\tau}(\cdot)} E_{\mathbf{h}}
        \left[\sum_{k=1}^K \mu_k
        \frac{\tau_k(\mathbf{h})}{h_k} (2^{r_k(\mathbf{h})}-1)\right]
        \\
        \mbox{subject to} \quad E_{\mathbf{h}} \left[\sum_{k=1}^K w_k \tau_k(\mathbf{h})
        r_k(\mathbf{h})\right]= \bar{R}, \quad \quad
        \forall \mathbf{h}\;\; \sum_{k=1}^K \tau_k(\mathbf{h}) =1.
    \end{cases}
\end{equation}
Using the Lagrange multiplier approach, we can decompose
(\ref{E:eq5}) into two sub-problems.

\begin{enumerate}

\item Given what we term the total rate-reward $R(\mathbf{h}):=
\sum_{k=1}^K w_k \tau_k(\mathbf{h}) r_k(\mathbf{h})$ assigned to
the $K$ users, we determine how to distribute $R(\mathbf{h})$
among users so that the total power cost in a {\it fixed} state
$\mathbf{h}$ is minimized. That is, we solve
\begin{equation}\label{E:eq5}
    \begin{cases}
        J(R(\mathbf{h})) := \min_{\mathbf{r}(\mathbf{h}),
        \boldsymbol{\tau}(\mathbf{h})} \sum_{k=1}^K \mu_k
        \frac{\tau_k(\mathbf{h})}{h_k} (2^{r_k(\mathbf{h})}-1)
        \\
        \mbox{subject to} \quad \sum_{k=1}^K w_k \tau_k(\mathbf{h})
        r_k(\mathbf{h}) = R(\mathbf{h}), \quad \quad
        \sum_{k=1}^K \tau_k(\mathbf{h}) =1.
    \end{cases}
\end{equation}
Upon defining the rate-reward for user $k$ as $R_k(\mathbf{h}):=
w_k r_k(\mathbf{h})$ and the corresponding
power-cost$\thicksim$rate-reward (C$\thicksim$R) function
\begin{equation}
f_k(x):=\frac{\mu_k}{h_k}( 2^{x/w_k}-1),
\end{equation}
we can rewrite (\ref{E:eq5}) as
\begin{equation}\label{E:JR}
    \begin{cases}
        J(R(\mathbf{h})) := \min_{\mathbf{r}(\mathbf{h}),
        \boldsymbol{\tau}(\mathbf{h})} \sum_{k=1}^K \tau_k(\mathbf{h})
        f_k (R_k(\mathbf{h}))
        \\
        \mbox{subject to} \quad \sum_{k=1}^K \tau_k(\mathbf{h})
        R_k(\mathbf{h}) = R(\mathbf{h}), \quad \quad
        \sum_{k=1}^K \tau_k(\mathbf{h}) =1.
    \end{cases}
\end{equation}

\item Having obtained $J(\cdot)$ in (\ref{E:JR}), we optimize the
allocation of $R(\mathbf{h})$ across the realizations of
$\mathbf{h}$, so that the total power cost averaged over all
fading states is minimized; that is
\begin{equation}\label{E:wf}
    \begin{cases}
        \min_{R(\mathbf{h})} E_{\mathbf{h}} [J(R(\mathbf{h}))] - \lambda E_{\mathbf{h}}
        [R(\mathbf{h})]\\
        \mbox{subject to} \quad E_{\mathbf{h}} [R(\mathbf{h})] = \bar{R}
    \end{cases}
\end{equation}
where $\lambda$ denotes the associated Lagrange multiplier.

\end{enumerate}

To gain insight, we first solve (\ref{E:JR}) and (\ref{E:wf}) for
two-users before generalizing to $K$-users.

\subsubsection{Two-User Case}

With $f_1(x)$ and $f_2(x)$ denoting the C$\thicksim$R functions
corresponding to users 1 and 2, we first establish following
lemma.

\begin{lemma}
{\it Supposing w.l.o.g. that $w_1<w_2$, it holds that:
\begin{enumerate}
\item If $\frac{\mu_1}{w_1h_1} \geq \frac{\mu_2}{w_2h_2}$, then
$f_2(x)<f_1(x)$, $\forall x > 0$.

\item If $\frac{\mu_1}{w_1h_1} < \frac{\mu_2}{w_2h_2}$, then
\begin{equation}
    \begin{cases}
        f_2(x) > f_1(x), & \mbox{when } 0 < x < v_0, \\
        f_2(x) < f_1(x), & \mbox{when } x > v_0;
    \end{cases}
\end{equation}
where $v_0$ is the unique solution to the equation
$f_2(x)=f_1(x)$.
\end{enumerate}}
\end{lemma}

\emph{Proof:} See Appendix A1. \hfill $\Box$

\begin{figure}
\centering \epsfig{file=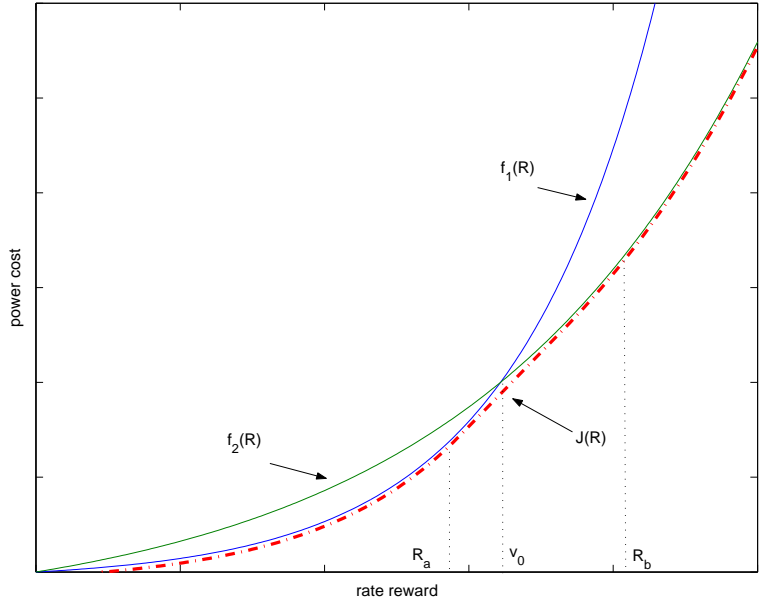, width=0.6\textwidth} \caption{
The functions $f_1(R)$, $f_2(R)$ and $J(R)$ when $w_1 < w_2$ and
$\frac{\mu_1}{w_1h_1} < \frac{\mu_2}{w_2h_2}$ (the dash-dotted
curve for $J(R)$ is slightly offset for easy visualization). }
\label{F:crf}
\end{figure}

When $w_1<w_2$, Lemma 1 asserts that if $\frac{\mu_1}{w_1h_1} \geq
\frac{\mu_2}{w_2h_2}$, the C$\thicksim$R curve $f_1(x)$ of user 1
stays always above $f_2(x)$. If $\frac{\mu_1}{w_1h_1} <
\frac{\mu_2}{w_2h_2}$, the two C$\thicksim$R curves cross each
other once at $v_0$, as shown in Fig.\ \ref{F:crf}; hence,
$\forall x \in (0, v_0)$, we have $f_2(x)>f_1(x)$; and for $x >
v_0$, $f_2(x)<f_1(x)$. Using Lemma 1, we can characterize
$J(R(\mathbf{h}))$ in (\ref{E:JR}) as follows.

\begin{lemma}
{\it For $K=2$ and $w_1<w_2$, the solution to (\ref{E:JR}) is:
\begin{enumerate}
\item If $\frac{\mu_1}{w_1h_1} \geq \frac{\mu_2}{w_2h_2}$, then
$J(R(\mathbf{h}))=f_2(R(\mathbf{h}))$, which is achieved by the
allocation $\tau_1^*(\mathbf{h})=0$, $r_1^*(\mathbf{h})=0$,
$\tau_2^*(\mathbf{h})=1$, and
$r_2^*(\mathbf{h})=R(\mathbf{h})/w_2$.

\item If $\frac{\mu_1}{w_1h_1} < \frac{\mu_2}{w_2h_2}$, then
\begin{equation}\label{E:eq10}
    J(R(\mathbf{h})) =
    \begin{cases}
        f_1(R(\mathbf{h})), & \mbox{if} \;\; 0 < R(\mathbf{h})
            \leq R_a(\mathbf{h}); \\
        f_2(R(\mathbf{h})), & \mbox{if} \;\; R(\mathbf{h})
            \geq R_b(\mathbf{h}); \\
        f_1(R_a(\mathbf{h}))+s_0(\mathbf{h}) [R(\mathbf{h})-
            R_a(\mathbf{h})], & \mbox{if} \;\;  R_a(\mathbf{h}) <
            R < R_b(\mathbf{h});
    \end{cases}
\end{equation}
where
\begin{equation}\label{E:Rabh}
    R_a(\mathbf{h}) = w_1 \log
    \left(\frac{s_0(\mathbf{h})w_1h_1}{\ln 2 \mu_1}\right) > 0, \quad
    \quad  R_b(\mathbf{h}) = w_2 \log
    \left(\frac{s_0(\mathbf{h})w_2h_2}{\ln 2 \mu_2}\right) > 0,
\end{equation}
and $s_0(\mathbf{h})$ is the solution of the equation
\begin{equation}\label{E:gx}
    g(x,\mathbf{h}):= x \left( w_2 \log
    \left(\frac{xw_2h_2}{\ln 2 \mu_2}\right) - w_1 \log
    \left(\frac{xw_1h_1}{\ln 2 \mu_1}\right) - \frac{w_2-w_1}{\ln 2}\right) +
    \frac{\mu_2}{h_2}-\frac{\mu_1}{h_1} =0,
\end{equation}
i.e., $g(s_0(\mathbf{h}),\mathbf{h})=0$. The minimum cost
$J(R(\mathbf{h}))$ is then achieved with these policies:

\begin{enumerate}
\item if  $0 < R(\mathbf{h}) \leq R_a(\mathbf{h})$, then
\begin{equation}\label{E:rt1}
    \begin{cases}
        \tau_1^*(\mathbf{h})=1, & r_1^*(\mathbf{h})= R(\mathbf{h})/ w_1, \\
        \tau_2^*(\mathbf{h})=0, & r_2^*(\mathbf{h})=0;
    \end{cases}
\end{equation}
\item if $R(\mathbf{h}) \geq R_b(\mathbf{h})$, then
\begin{equation}
    \begin{cases}
        \tau_1^*(\mathbf{h})=0, & r_1^*(\mathbf{h})= 0, \\
        \tau_2^*(\mathbf{h})=1, & r_2^*(\mathbf{h})=R(\mathbf{h})/ w_2;
    \end{cases}
\end{equation}
\item if $R_a(\mathbf{h}) < R(\mathbf{h}) < R_b(\mathbf{h})$, then
\begin{equation}\label{E:rt3}
    \begin{cases}
        \tau_1^*(\mathbf{h})=\frac{R(\mathbf{h})-R_a(\mathbf{h})}{R_b(\mathbf{h})-R_a(\mathbf{h})},
        & r_1^*(\mathbf{h})= R_a(\mathbf{h})/ w_1, \\
        \tau_2^*(\mathbf{h})=1-\tau_1(\mathbf{h}), & r_2^*(\mathbf{h})=R_b(\mathbf{h})/ w_2.
    \end{cases}
\end{equation}
\end{enumerate}
\end{enumerate}}
\end{lemma}

\emph{Proof:} See Appendix A2. \hfill $\Box$

Lemma 2 specifies the optimal power cost curve $J(R(\mathbf{h}))$
for each fading state $\mathbf{h}$ when $w_1 < w_2$. Specifically,
if $\frac{\mu_1}{w_1h_1} \geq \frac{\mu_2}{w_2h_2}$,
$J(R(\mathbf{h}))$ is simply $f_2(R(\mathbf{h}))$; otherwise,
$J(R(\mathbf{h}))$ comprises part of the $f_1(R(\mathbf{h}))$
curve, the tangent line, and part of the $f_2(R(\mathbf{h}))$
curve, as indicated in Fig.\ \ref{F:crf}. Having obtained
$J(R(\mathbf{h}))$, we now solve (\ref{E:wf}) to obtain the
optimal resource allocation policies.

\begin{theorem}
{\it For $K=2$ and $w_1 < w_2$, the optimal rate and time
allocation policies with respect to the minimization problem
(\ref{E:wsum}) are as follows:

\begin{enumerate}
\item If $\frac{\mu_1}{w_1h_1} \geq \frac{\mu_2}{w_2h_2}$, then
\begin{equation}\label{E:2U1}
    \begin{cases}
        \tau_1^*(\mathbf{h})=0, & r_1^*(\mathbf{h})= 0, \\
        \tau_2^*(\mathbf{h})=1, & r_2^*(\mathbf{h})=\left[ \log \lambda^* -
        \log \left( \frac{\ln 2\mu_2}{w_2h_2}\right)\right]_+.
    \end{cases}
\end{equation}

\item If $\frac{\mu_1}{w_1h_1} < \frac{\mu_2}{w_2h_2}$, and
$\xi:=\left( \frac{((w_1h_1)/(\ln 2\mu_1))^{w_1}}{((w_2h_2)/(\ln
2\mu_2))^{w_2}} \right)^{\frac{1}{w_2-w_1}}$,
\begin{enumerate}
\item if $\lambda^* \leq \xi$ or if $\lambda^* > \xi$ and
$g(\lambda^*,\mathbf{h}) < 0$, then
\begin{equation}\label{E:2U2}
    \begin{cases}
        \tau_1^*(\mathbf{h})=1, & r_1^*(\mathbf{h})= \left[ \log \lambda^* -
        \log \left( \frac{\ln 2\mu_1}{w_1h_1}\right)\right]_+, \\
        \tau_2^*(\mathbf{h})=0, & r_2^*(\mathbf{h})=0;
    \end{cases}
\end{equation}

\item if $\lambda^* > \xi$ and $g(\lambda^*,\mathbf{h}) > 0$, then
\begin{equation}\label{E:2U3}
    \begin{cases}
        \tau_1^*(\mathbf{h})=0, & r_1^*(\mathbf{h})= 0, \\
        \tau_2^*(\mathbf{h})=1, & r_2^*(\mathbf{h})=\left[ \log \lambda^* -
        \log \left( \frac{\ln 2\mu_2}{w_2h_2}\right)\right]_+;
    \end{cases}
\end{equation}

\item if $\lambda^* > \xi$ and $g(\lambda^*,\mathbf{h}) = 0$, then
for an arbitrary $\tau_0^* \in [0,1]$,
\begin{equation}\label{E:2U4}
    \begin{cases}
        \tau_1^*(\mathbf{h})=\tau_0^*, & r_1^*(\mathbf{h})= \left[ \log \lambda^* -
        \log \left( \frac{\ln 2\mu_1}{w_1h_1}\right)\right]_+, \\
        \tau_2^*(\mathbf{h})=1-\tau_0^*,
        & r_2^*(\mathbf{h})=\left[ \log \lambda^* -
        \log \left( \frac{\ln 2\mu_2}{w_2h_2}\right)\right]_+.
    \end{cases}
\end{equation}
\end{enumerate}
\end{enumerate}
Function $g(x,\mathbf{h})$ is given by (\ref{E:gx}), and
$\lambda^*$, $\tau_0^*$ are obtained numerically by satisfying the
weighted sum-rate constraint $E_\mathbf{h}[\sum_{k=1}^2 w_k
\tau_k^*(\mathbf{h}) r_k^*(\mathbf{h})] = \bar{R}$.}
\end{theorem}

\emph{Proof:} See Appendix B. \hfill $\Box$

Note that depending on $\lambda^*$, water-filling in
(\ref{E:2U1})-(\ref{E:2U4}) may result in zero transmission rate
for the user which has been assigned the entire or part of the
block. Therefore, for some fading states where the channel is
really bad, both users should defer. Comparing (\ref{E:JR}) and
(\ref{E:wf}) with \cite[eqs. (11), (13)]{Li&Goldsmith2001a}, we
find that our {\it power minimization} yields similar
``opportunistic'' policies as the {\it rate maximization} in
\cite{Li&Goldsmith2001a}.

Since $\tau_0^*$ can take any arbitrary value between 0 and 1, the
solution in (\ref{E:2U4}) is not unique. However, under the
assumption of continuous joint fading distribution density, the
probability of $g(\lambda^*,\mathbf{h}) = 0$ is zero, and
therefore case c) is an event of measure zero. Thus, after setting
$\tau_0^*$ to an arbitrary value in [0,1], $\lambda^*$ can be
uniquely determined by an one-dimensional, e.g., bi-sectional,
search. From Theorem 1, it is clear that in order to achieve
energy efficiency over TDMA fading channels, most of the time we
should allow one user to transmit per block. This also holds true
in rate maximization for TDMA fading
channels~\cite{Li&Goldsmith2001a}, even though the resultant time
and rate allocation fractions are different.

To extend our two-user results to $K>2$ users, we will need
definition of the {\it convex envelope}.

\begin{definition}
{\it The convex envelope $f^c(x)$ of a function $f(x)$ is the
solution to the optimization problem
\begin{equation}
    f^c(x) = \max_{a,b}\;\; ax+b, \quad \quad \mbox{subject to} \quad
    ax+b \leq f(x), \; \forall x.
\end{equation}
Namely, $f^c(x)$ is the boundary surface of the convex hull of the
function's epigraph~\cite{convex}.}
\end{definition}

Using the definition $\bar{f}(x):=\min_{1 \leq k \leq K} f_k(x)$,
we can verify the following property:

\begin{proposition}
{\it The optimal C$\thicksim$R function $J(R(\mathbf{h}))$ in
(\ref{E:JR}) is the convex envelope of $\bar{f}(R(\mathbf{h}))$ in
the two-user case; i.e.,
$J(R(\mathbf{h}))=\bar{f}^c(R(\mathbf{h}))$ as determined by Lemma
2 and Definition 2.}
\end{proposition}

\emph{Proof:} If $w_1<w_2$ and  $\frac{\mu_1}{w_1h_1} \geq
\frac{\mu_2}{w_2h_2}$, then
$J(R(\mathbf{h}))=f_2(R(\mathbf{h}))=\bar{f}(R(\mathbf{h}))$. It
is trivial to show $J(R(\mathbf{h}))=\bar{f}^c(R(\mathbf{h}))$. If
$w_1<w_2$ and  $\frac{\mu_1}{w_1h_1} < \frac{\mu_2}{w_2h_2}$, then
it is easy to show that $J(R(\mathbf{h}))$ is convex since it has
non-decreasing first derivatives for all $R(\mathbf{h}) \geq 0$.
If the convex envelope $\bar{f}^c(R(\mathbf{h}))$ were not given
by $J(R(\mathbf{h}))$, then $\bar{f}^c(R(\mathbf{h}))$ would be
strictly greater than $J(R(\mathbf{h}))$ for some $R(\mathbf{h})$.
Since the first and the third branches of $J(R(\mathbf{h}))$ in
(\ref{E:eq10}) are exactly $\bar{f}(R(\mathbf{h}))$, for $0 <
R(\mathbf{h}) \leq R_a(\mathbf{h})$ and $R(\mathbf{h}) \geq
R_b(\mathbf{h})$, we must have
$\bar{f}^c(R(\mathbf{h}))=J(R(\mathbf{h}))$. Therefore,
$\bar{f}^c(R(\mathbf{h}))$ can only be greater than
$J(R(\mathbf{h}))$ for $R(\mathbf{h}) \in (R_a(\mathbf{h}),
R_b(\mathbf{h}))$. But since $\bar{f}^c(R(\mathbf{h}))$ is convex,
Jensen's inequality implies that its value for any
$R_a(\mathbf{h}) < R(\mathbf{h}) < R_b(\mathbf{h})$ can not be
greater than the value given by the line segment connecting
$\bar{f}(R_a(\mathbf{h}))$ with $\bar{f}(R_b(\mathbf{h}))$. This
leads to a contradiction, and thus
$J(R(\mathbf{h}))=\bar{f}^c(R(\mathbf{h}))$. \hfill $\Box$

\subsubsection{$K$-Use Case}

Generalizing Proposition 1 to $K>2$ users, we can show that:
\begin{theorem}
{\it For a $K$-user TDMA block fading channel, the optimal
C$\thicksim$R function $J(R(\mathbf{h}))$ at each fading state
$\mathbf{h}$ is the convex envelope of
$\bar{f}(R(\mathbf{h})):=\min_{1 \leq k \leq K}
f_k(R(\mathbf{h}))$, and the optimality is achieved by allowing at
most two users to transmit per time block.}
\end{theorem}

\emph{Proof:} See Appendix C. \hfill $\Box$

Although Theorem 2 asserts achievability of the optimal policies,
it does not provide algorithms realizing these optimal allocation
strategies. The latter are challenging since obtaining convex
envelopes is generally difficult. As pointed out
in~\cite{Kumaran2005}, the rate-maximizing resource allocation for
TD broadcast fading channels in~\cite{Li&Goldsmith2001a} is
obtained by water-filling across realizations of concave
envelopes. Our energy-efficient resource allocation policies under
a weighted sum average rate constraint for TDMA fading channels
are obtained via water-filling across realizations of the convex
envelopes $J(R(\mathbf{h}))$. But in order to determine
$J(R(\mathbf{h}))$, we must generalize Lemma 2 to the $K$-user
case. This is accomplished with the following algorithm.

\begin{algorithm}
{\it initialization:} Let $\mathbf{u}:=[u_1, \ldots, u_{K_0}]^T$
(initially set equal to the empty set $\phi$) denote the set of
active users during a given block, $\mathbf{s}:=[s_1, \ldots,
s_{K_0}]^T$ the set of slopes of (possibly multiple) tangent lines
common to the active C$\thicksim$R curves (also initially set to
$\phi$), and let the iteration index be $m=1$.

\begin{enumerate}\renewcommand{\labelenumi}{{\it S\arabic{enumi})}}

\item Remove the C$\thicksim$R function of user $k$ from
$J(R(\mathbf{h}))$ if $\exists i \neq k$ such that $w_k \leq w_i$
and $\frac{\mu_k}{w_kh_k} \geq \frac{\mu_i}{w_ih_i}$, because in
this case $\forall x>0$, $f_k(x)<f_i(x)$ and $f_k(x)$ will not
appear in the expression of $J(R(\mathbf{h}))$ for the reasons we
detailed in Lemma 2. Let $K_r$ be the number of users remaining
after such a successive elimination of their C$\thicksim$R
functions from $J(R(\mathbf{h}))$, and define the permutation
$\pi(\cdot)$ such that
$w_{\pi(1)}<w_{\pi(2)}<\cdots<w_{\pi(K_r)}$. Then through
successive pairwise comparisons and user C$\thicksim$R function
removals, we can ensure that
$\frac{\mu_{\pi(1)}}{w_{\pi(1)}h_{\pi(1)}} <
\frac{\mu_{\pi(2)}}{w_{\pi(2)}h_{\pi(2)}} < \cdots <
\frac{\mu_{\pi(K_r)}}{w_{\pi(K_r)}h_{\pi(K_r)}}$.

\item Let the $m$th element of $\mathbf{u}$ be $u_m=\pi(1)$.  If
$K_r \geq 2$, go to {\it Step 3}. If $K_r<2$, all users with
C$\thicksim$R functions not appearing in $J(R(\mathbf{h}))$ have
been removed. Then set the number of active users be $K_0=m$,
$s_m=\infty$, and stop.

\item For $1 \leq i \leq K_r-1$, define
\begin{eqnarray}
    \nonumber
    g_i(x,\mathbf{h})& :=  & x \left( w_{\pi(i+1)} \log
    \left(\frac{xw_{\pi(i+1)}h_{\pi(i+1)}}{\ln 2\mu_{\pi(i+1)}}\right) - w_{\pi(1)} \log
    \left(\frac{xw_{\pi(1)}h_{\pi(1)}}{\ln 2\mu_{\pi(1)}}\right) \right) \\
    & & - x\left(\frac{w_{\pi(i+1)}-w_{\pi(1)}}{\ln 2}\right) +
    \frac{\mu_{\pi(i+1)}}{h_{\pi(i+1)}}-\frac{\mu_{\pi(1)}}{h_{\pi(1)}},
\end{eqnarray}
and find the $\xi_i$ satisfying $g_i(\xi_i,\mathbf{h})=0$,
$\log\left(\frac{xw_{\pi(i+1)}h_{\pi(i+1)}}{\ln
2\mu_{\pi(i+1)}}\right) >0$ and
$\log\left(\frac{xw_{\pi(1)}h_{\pi(1)}}{\ln
2\mu_{\pi(1)}}\right)>0$. Let also $s_m=\min_i \{\xi_i\}$, and
$i^*=\arg \; \min_i \{\xi_i\}$. Remove C$\thicksim$R functions of
users $\pi(k)$ for which $1 \leq k \leq i^*$. Increase $m$ by 1
and return to {\it S1)}.

\end{enumerate}
\end{algorithm}

Using Algorithm 1, we obtain $\mathbf{u}$ and $\mathbf{s}$. If
$K_0=1$, $J(R(\mathbf{h}))$ is simply $f_{u_1}(\cdot)$; if
$K_0>1$, then $J(R(\mathbf{h}))$ comprises pieces of the curves
$\left\{f_{u_m}(\cdot)\right\}_{m=1}^{K_0}$ as well as the common
tangent line segments between $f_{u_m}(\cdot)$ and
$f_{u_{m+1}}(\cdot)$, $1 \leq m \leq K_0-1$, the slopes of which
are $\left\{s_m\right\}_{m=1}^{K_0-1}$. By denoting
$R_{a_0}(\mathbf{h})=R_{b_0}(\mathbf{h})=0$,
$R_{a_{K_0}}(\mathbf{h})=R_{b_{K_0}}(\mathbf{h})=\infty$, and
letting $R_{a_m}(\mathbf{h})$ and $R_{b_m}(\mathbf{h})$ be the
points with equal first derivatives
$f_{u_m}^{(1)}(R_{a_m}(\mathbf{h}))=f_{u_{m+1}}^{(1)}(R_{b_m}(\mathbf{h}))=s_m$,
we can write
\begin{equation}
    J(R(\mathbf{h})) =
    \begin{cases}
        f_{u_m}(R(\mathbf{h})), & \mbox{if} \;\; R_{b_{m-1}}(\mathbf{h}) \leq R(\mathbf{h})
            \leq R_{a_m}(\mathbf{h}); \\
        f_{u_m}(R_{a_m}(\mathbf{h}))+s_m [R(\mathbf{h})-
            R_{a_m}(\mathbf{h})], & \mbox{if} \;\;  R_{a_m}(\mathbf{h})
            \leq R < R_{b_m}(\mathbf{h}).
    \end{cases}
\end{equation}

Arguing as in the proof of Proposition 1, it follows readily that
$J(R(\mathbf{h}))$ is the convex envelope of
$\bar{f}(R(\mathbf{h}))$. Once having $J(R(\mathbf{h}))$, we will
implement a water-filling strategy to obtain the energy-efficient
resource allocation across realizations of $\mathbf{h}$. First,
for any user $k \notin \mathbf{u}$, we let
$\tau_k^*(\mathbf{h})=0$ and $r_k^*(\mathbf{h})= 0$. The rate and
time allocation for the remaining $K_0$ users is given as follows.

\begin{algorithm}
\begin{enumerate}
\item If $K_0=1$, then
\begin{equation}\label{E:eq22}
    \tau_{u_1}^*(\mathbf{h})=1, \quad \quad r_{u_1}^*(\mathbf{h})= \left[ \log \lambda^* -
        \log \left( \frac{\ln 2\mu_{u_1}}{w_{u_1}h_{u_1}}\right)\right]_+.
\end{equation}

\item If $K_0>1$, set $s_0=0$. Since
$0=s_0<s_1<\cdots<s_{K_0}=\infty$, for the given $\lambda^*$,
there exists $j \in \{1,2,\ldots, K_0\}$ such that $s_{j-1} <
\lambda^* < s_j$ or $\lambda^*=s_j$.

\begin{enumerate}
\item If $\exists j$ such that $s_{j-1} < \lambda^* < s_j$, we
know $J(R(\mathbf{h}))=f_{u_j}(R(\mathbf{h}))$. In this case, we
set
\begin{equation}
    \tau_{u_j}^*(\mathbf{h})=1, \quad \quad r_{u_j}^*(\mathbf{h})= \left[ \log \lambda^* -
        \log \left(
        \frac{\ln 2\mu_{u_j}}{w_{u_j}h_{u_j}}\right)\right]_+;
\end{equation}
and $\forall i \neq j$, $\tau_{u_i}^*(\mathbf{h})=0$ and
$r_{u_i}^*(\mathbf{h})=0$.

\item If $\exists j$ such that $\lambda^*=s_j$, then
$J(R(\mathbf{h}))=f_{u_j}(R_{a_j}(\mathbf{h}))+s_j [R(\mathbf{h})-
R_{a_j}(\mathbf{h})]$. As in the two-user case, we set
\begin{equation}\label{E:eq24}
    \begin{cases}
        \tau_{u_j}^*(\mathbf{h})=\tau_0^*, & r_{u_j}^*(\mathbf{h})=  \left[ \log \lambda^* -
        \log \left( \frac{\ln 2\mu_{u_j}}{w_{u_j}h_{u_j}}\right)\right]_+, \\
        \tau_{u_{j+1}}^*(\mathbf{h})=1-\tau_0^*,
        & r_{u_{j+1}}^*(\mathbf{h})= \left[ \log \lambda^* -
        \log \left(
        \frac{\ln 2\mu_{u_{j+1}}}{w_{u_{j+1}}h_{u_{j+1}}}\right)\right]_+;
    \end{cases}
\end{equation}
and $\forall i \neq j, j+1$, $\tau_{u_i}^*(\mathbf{h})=0$ and
$r_{u_i}^*(\mathbf{h})=0$.
\end{enumerate}
\end{enumerate}

In eqs. (\ref{E:eq22})-(\ref{E:eq24}), $\lambda^*$ and $\tau_0^*$
are chosen to satisfy the weighted sum-rate constraint
\begin{equation}
    E_\mathbf{h}\left[\sum_{k=1}^K w_k \tau_k^*(\mathbf{h})
    r_k^*(\mathbf{h})\right] = \bar{R}.
\end{equation}
\end{algorithm}

Under the assumption of continuous joint fading distribution
density, the value of $\tau_0^*$ does not affect the weighted
sum-rate constraint, and $\lambda^*$ can be uniquely determined by
an one-dimensional search, as in the two-user case. To achieve
energy efficiency, we should only allow at most two users (and
most of the time only one user) to transmit per time block in the
$K$-user case. Again in (\ref{E:eq22})-(\ref{E:eq24}),
water-filling may result in zero transmission rate for the user
which has been assigned the entire or part of the time block. For
some fading states, when all channels are in deep fading, all
users should defer. Also note that Algorithms 1 and 2 are dual to
those in~\cite{Li&Goldsmith2001a} for power minimization under an
average sum-rate constraint.

\subsection{Quantized CSI and Finite AMC Modes}\label{S:AMC}

In this section, we provide a novel formulation and solve the
energy minimization problem under a weighted sum average rate
constraint for the {\it finite-AMC-mode} case. In practice, a user
may not be able to support an infinite number of codebooks.
Moreover, the codewords in use may not be capacity-achieving. It
is thus worth investigating energy-efficient resource allocation
for practical systems where each user can only support a finite
number of AMC modes. Notice that since transmitters can transmit
with a finite number of AMC modes, only {\it quantized} CSI could
be fed back from the access point to the transmitters suffices.

For user $k \in [1, K]$, an AMC mode corresponds to a rate-power
pair $(\rho_{k,l}, p_{k,l})$, $l=1,\ldots,M_k$, where $M_k$
denotes the number of AMC modes. A pair $(\rho_{k,l}, p_{k,l})$
indicates that for transmission rate $\rho_{k,l}$ provided by the
$l$th AMC mode, the minimum received power required is $p_{k,l}$.
Notice that the minimum power $p_{k,l}$ may not be given by
$2^{\rho_{k,l}}-1$ as in the capacity-achieving case, and some
extra power may be required in practice. Also, the rate
$\rho_{k,l}$ is maintained with a prescribed symbol error
probability (SEP), and $p_{k,l}$ is the corresponding minimum
received power under the SEP constraint. For this reason, we need
to implicitly include the SEP constraints in our optimization.
Although the $k$th user only supports $M_k$ AMC modes, this user
can still support through time-sharing continuous rates up to a
maximum value determined by the highest-rate AMC mode
$\rho_{k,M_k}$.

By setting $\rho_{k,0}=0$ and $p_{k,0}=0$ and letting
$\gamma_{k,l}:=(p_{k,l}-p_{k,l-1})/(\rho_{k,l}-\rho_{k,l-1})$, we
define the {\it piece-wise linear} function relating transmit
power with rate as
\begin{equation}\label{E:upsilon}
    \Upsilon_k (r_k(\mathbf{h})) =
    \begin{cases}
        p_{k,l-1}/h_k + \gamma_{k,l}
        (r_k(\mathbf{h})-\rho_{k,l-1})/h_k, & \rho_{k,l-1} \leq r_k(\mathbf{h}) \leq
        \rho_{k,l}, \;\; l \in [1, M_k]; \\
        \infty, & r_k(\mathbf{h}) > \rho_{k,M_k}.
    \end{cases}
\end{equation}
Notice that in order to support rate $\rho_{k,l}$ with channel
coefficient $h_k$, the required transmit power is scaled as
$p_{k,l}/h_k$. For practical modulation-coding schemes with e.g.,
$M$-QAM constellations and error-control codes, $\Upsilon_k
(r_k(\mathbf{h}))$ is guaranteed to be
convex~\cite{Uysal-Biyikoglu02net}. Using (\ref{E:upsilon}) to
replace the power-rate relationship implied by Shannon's capacity
formula, we can define a power region as [c.f. (\ref{E:power}),
(\ref{E:PTD})]
\begin{equation}\label{E:region2}
    {\cal P}'(\bar{R}) = \bigcup_{(\mathbf{r}(\cdot),\boldsymbol{\tau}(\cdot)) \in {\cal
    F}_{\mathbf{w}}} {\cal P}_{TD}' (\mathbf{r}(\cdot),\boldsymbol{\tau}(\cdot)),
\end{equation}
where
\begin{equation}\label{E:wsum2}
    {\cal P}_{TD}' (\mathbf{r}(\cdot),\boldsymbol{\tau}(\cdot)) = \left\{ \mathbf{\bar{p}}:
    \bar{P}_k
    \geq E_\mathbf{h}\left[
    \tau_k(\mathbf{h})\Upsilon_k
    (r_k(\mathbf{h}))
    \right], \quad \quad 1 \leq k \leq K
    \right\}.
\end{equation}
It is easy to show that the $K$-dimensional ${\cal P}'(\bar{R})$
is feasible and convex. The optimization problem thus becomes
\begin{equation}\label{E:optimal2}
    \min_{\mathbf{\bar{p}}} \boldsymbol{\mu}^T \mathbf{\bar{p}}, \quad \quad
    \mbox{subject to} \quad  \mathbf{\bar{p}} \in {\cal
    P}'(\bar{R}).
\end{equation}
We can rewrite (\ref{E:optimal2}) as
\begin{equation}\label{E:eq31}
    \begin{cases}
        \min_{\mathbf{r}(\cdot),\boldsymbol{\tau}(\cdot)} E_\mathbf{h}\left[
        \sum_{k=1}^K \mu_k \tau_k(\mathbf{h})\Upsilon_k(r_k(\mathbf{h}))\right]
        \\
        \mbox{subject to} \quad E_{\mathbf{h}} \left[\sum_{k=1}^K w_k \tau_k(\mathbf{h})
        r_k(\mathbf{h})\right]= \bar{R}, \quad \quad
        \forall \mathbf{h}\;\; \sum_{k=1}^K \tau_k(\mathbf{h}) =1.
    \end{cases}
\end{equation}
As in the infinite-codebook case, we can still decompose
(\ref{E:eq31}) into two sub-problems. That is, we first optimize
per fading realization, and then apply water-filling across
$J(R(\mathbf{h}))$ realizations to obtain the optimal resource
allocation policies.

Recalling the weighted rate-reward $R_k(\mathbf{h})=w_k
r_k(\mathbf{h})$, the C$\thicksim$R function corresponding to the
$k$th user is now $f_k(R_k(\mathbf{h})) :=\mu_k \Upsilon_k
(R_k(\mathbf{h})/w_k)$ which is a piece-wise linear curve through
the points $\{(w_k\rho_{k,l}, \mu_k p_{k,l}/h_k)\}_{l=1}^{M_k}$.
Since the C$\thicksim$R functions are piece-wise linear, the
convex envelope $J(R(\mathbf{h}))$ of their $\min_k \;
f_k(R(\mathbf{h}))$ can be obtained by simply comparing the slopes
of a finite number of straight line segments. To this end, we
implement the following algorithm:

\begin{algorithm}
{\it initialization:} Define the set of points $\bar{\mathbf{m}}_k
:= [ (w_k\rho_{k,l},
\mu_k p_{k,l}/h_k), \;\; l=1,\ldots,M_k ]^T$, $\forall k \in [1, K]$. 
Start with the set of slopes $\mathbf{s}:=[s_0, \ldots,
s_{m-1}]^T$, set of rate rewards $\mathbf{r}:=[R_0, \ldots,
R_{m-1}]^T$, set of power costs $\mathbf{c}:=[C_0, \ldots,
C_{m-1}]^T$ with $s_0=R_0=C_0=0$, and let the iteration index be
$m=1$.

\begin{enumerate}\renewcommand{\labelenumi}{{\it S\arabic{enumi})}}

\item Consider the point $(x_0, y_0)=(R_{m-1}, C_{m-1})$. For
$k=1,\ldots, K$, if $\bar{\mathbf{m}}_k \neq \phi$, let the $i$th
element in $\bar{\mathbf{m}}_k$ be
$(w_k\rho_{k,\bar{\mathbf{m}}_k(i)}, \mu_k
p_{k,\bar{\mathbf{m}}_k(i)}/h_k)$. Also, set $(x_k ,
y_k)=\left(w_k\rho_{k,\bar{\mathbf{m}}_k(1)},p_{k,\bar{\mathbf{m}}_k(1)}/h_k\right)$.

\item Let $s_m= \min_{k:\;\bar{\mathbf{m}}_k \neq \phi} \;
\frac{y_k-y_0}{x_k-x_0}$ and $k^*= \arg \;
\min_{k:\;\bar{\mathbf{m}}_k \neq \phi} \;
\frac{y_k-y_0}{x_k-x_0}$. Set $R_m=x_{k^*}$ and $C_m=y_{k^*}$.

\begin{enumerate}
\item For each $k \neq k^*$ and $\bar{\mathbf{m}}_k \neq \phi$, $k
\in [1, K]$, let $i^* = \max_i \; \{i:
w_k\rho_{k,\bar{\mathbf{m}}_k(i)} \leq
w_{k^*}\rho_{k^*,\bar{\mathbf{m}}_{k^*}(1)}\}$. If $i^* \neq 0$,
then remove all $(w_k\rho_{k,\bar{\mathbf{m}}_k(i)}, \mu_k
p_{k,\bar{\mathbf{m}}_k(i)}/h_k)$, $i \in [1, i^*]$, from
$\bar{\mathbf{m}}_k$, and reduce $M_k$ by $i^*$. Note that in the
next iteration, the original
$(w_k\rho_{k,\bar{\mathbf{m}}_k(i^*+1)},$ \linebreak $\mu_k
p_{k,\bar{\mathbf{m}}_k(i^*+1)}/h_k)$ becomes the first element
$(w_k\rho_{k,\bar{\mathbf{m}}_k(1)}, \mu_k
q_{k,\bar{\mathbf{m}}_k(1)}/h_k)$ of $\bar{\mathbf{m}}_k$.

\item Remove $(w_{k^*}\rho_{k^*,\bar{\mathbf{m}}_{k^*}(1)},
\mu_{k^*} p_{k^*,\bar{\mathbf{m}}_{k^*}(1)}/h_{k^*})$ from
$\bar{\mathbf{m}}_{k^*}$ and reduce $M_{k^*}$ by 1.
\end{enumerate}

If $\bar{\mathbf{m}}_k = \phi$ (i.e., $M_k=0$), $\forall k \in
[1,K]$, set $K_0=m$ and stop. Otherwise, increase $m$ by 1 and go
to {\it S1)}. Notice that $K_0$ is the number of corner points of
the wanted convex envelope $J(R(\mathbf{h}))$.
\end{enumerate}
\end{algorithm}

\begin{figure}
\centering \epsfig{file=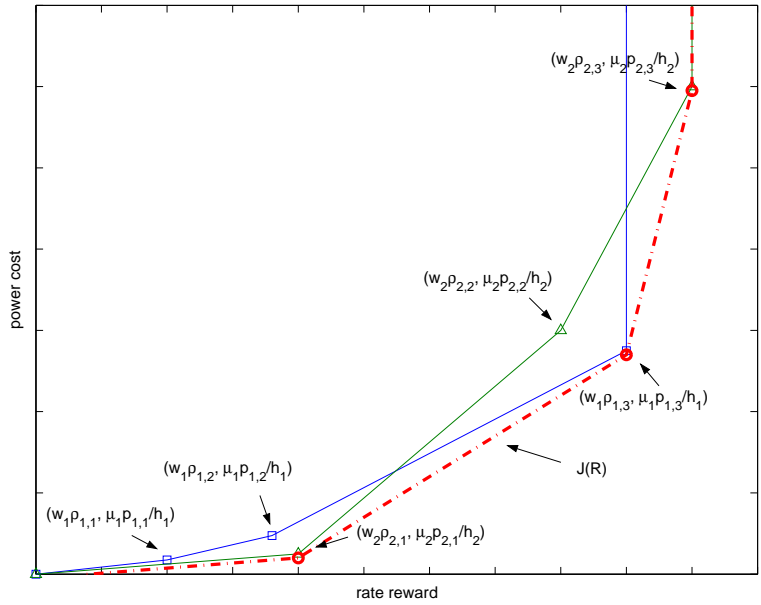, width=0.6\textwidth} \caption{
Convex envelope $J(R)$ in the finite-AMC-mode case (the
dash-dotted curve for $J(R)$ is slightly offset for easy
visualization). } \label{F:dcrf}
\end{figure}

Having obtained $\mathbf{s}$, $\mathbf{r}$ and $\mathbf{c}$, we
can express $J(R(\mathbf{h}))$ as
\begin{equation}\label{E:eq32}
    J(R(\mathbf{h}))=
    \begin{cases}
        C_{m-1}+s_m (R(\mathbf{h})-R_m), & R_{m-1} \leq
        R(\mathbf{h}) \leq R_m, \;\; m \in [1, K_0]; \\
        \infty, & R(\mathbf{h}) > R_{K_0}.
    \end{cases}
\end{equation}
An illustration example for Algorithm 3 and the resultant
$J(R(\mathbf{h}))$ is shown in Fig.\ \ref{F:dcrf}, where each of
the two users can support three AMC modes. In {\it S1} of
Algorithm 3, we first compare the slopes $\mu_1
p_{1,1}/(h_1w_1\rho_{1,1})$ and $\mu_2 p_{2,1}/(h_2w_2\rho_{2,1})$
and find $\mu_2 p_{2,1}/(h_2w_2\rho_{2,1}) < \mu_1
p_{1,1}/(h_1w_1\rho_{1,1})$. Therefore, we let $R_1 =
w_2\rho_{2,1}$ and $C_1 = \mu_2 p_{2,1}/h_2$. Since both
$w_1\rho_{1,1}$ and $w_1\rho_{1,2}$ are less than $w_2\rho_{2,1}$,
we remove the first two AMC modes of user 1 from
$\bar{\mathbf{m}}_1$ in {\it S2} a); whereas we remove the first
AMC mode of user 2 from $\bar{\mathbf{m}}_2$ in {\it S2} b). In
{\it S1} of the next iteration, we compare the slope between
points $(w_1\rho_{1,3}, \mu_1 p_{1,3}/h_1)$ and $(w_2\rho_{2,1},
\mu_2 p_{2,1}/h_2)$, with the slope between points
$(w_2\rho_{2,2}, \mu_2 p_{2,2}/h_2)$ and $(w_2\rho_{2,1}, \mu_2
p_{2,1}/h_2)$. In this case, we should set $R_2 = w_1\rho_{1,3}$
and $C_2 = \mu_1 p_{1,3}/h_1$. In {\it S2}, we remove
$(w_2\rho_{2,2}, \mu_2 p_{2,2}/h_2)$ from $\bar{\mathbf{m}}_2$,
and remove $(w_1\rho_{1,3}, \mu_1 p_{1,3}/h_1)$ from
$\bar{\mathbf{m}}_1$. In the last iteration, we obtain $R_3 =
w_2\rho_{2,3}$ and $C_3 = \mu_2 p_{2,3}/h_2$, and
$J(R(\mathbf{h}))$ is determined.

Having determined $J(R(\mathbf{h}))$ as in (\ref{E:eq32}), we
implement water-filling across realizations of $J(R(\mathbf{h}))$
to derive the optimal resource allocation policies. Different from
the infinite-codebook case where all $J(R(\mathbf{h}))$ have
continuous slopes, here $J(R(\mathbf{h}))$ are piecewise-linear.
Therefore, water-filling should take into account the finite
number of slopes of $J(R(\mathbf{h}))$. A somewhat related problem
was dealt with in~\cite{Li&Goldsmith2001b}, where water-filling
over some piecewise-linear concave functions was used to determine
the boundary surface of the outage probability region. But
energy-efficient resource allocation policies for finite-AMC-modes
were not considered in~\cite{Li&Goldsmith2001a} and
\cite{Li&Goldsmith2001b}.

Recall that $K_0$ and all entries of $\mathbf{s}$, $\mathbf{r}$
and $\mathbf{c}$ are functions of $\mathbf{h}$. Since every point
of the convex envelope $J(R(\mathbf{h}))$ can be achieved by
time-sharing between points $(R_m(\mathbf{h}), C_m(\mathbf{h}))$,
finding the optimal resource allocation strategies is equivalent
to solving the following minimization problem:
\begin{equation}\label{E:optimal3}
    \begin{cases}
        \min_{\mathbf{\tilde{\tau}}(\mathbf{h})} \; E_{\mathbf{h}}
        \left[\sum_{m=1}^{K_0} \tilde{\tau}_m(\mathbf{h})
        C_m(\mathbf{h})\right] \\
        \mbox{subject to} \quad \quad E_{\mathbf{h}}
        \left[\sum_{m=1}^{K_0} \tilde{\tau}_m(\mathbf{h})
        R_m(\mathbf{h})\right] \geq \bar{R}, \quad
        \sum_{m=1}^{K_0} \tilde{\tau}_m(\mathbf{h}) = 1.
    \end{cases}
\end{equation}

\begin{theorem}
{\it If the optimization (\ref{E:optimal3}) is feasible, by
setting $s_{K_0+1}(\mathbf{h})=\infty$, then $\forall \mathbf{h}$,
we have the optimal solution
$\left\{\tilde{\tau}_m^*(\mathbf{h})\right\}_{m=1}^{K_0}$, and
thus the optimal allocation policies $\tau_k^*(\mathbf{h})$ and
$r_k^*(\mathbf{h})$ ($1 \leq k \leq K$) for the original problem
(\ref{E:optimal2}) as follows:

\begin{enumerate}
\item If $\lambda^* < s_1(\mathbf{h})$, then
$\tilde{\tau}_m^*(\mathbf{h})=0$, $\forall m=1,\ldots, K_0$;
consequently, $\tau_k^*(\mathbf{h})=0$ and $r_k^*(\mathbf{h})=0$,
$\forall k=1,\ldots, K$.

\item If $\exists m^* \in \{1,2,\ldots,K_0\}$ so that
$s_{m^*}(\mathbf{h}) < \lambda^* < s_{m^*+1}(\mathbf{h})$, then
$\tilde{\tau}_{m^*}^*(\mathbf{h})=1$, and
$\tilde{\tau}_m^*(\mathbf{h})=0$, $\forall m \neq m^*$, $1 \leq m
\leq K_0$. This implies that if $(R_{m^*}(\mathbf{h}),
C_{m^*}(\mathbf{h}))$ belongs to user $k^*$, then
\begin{equation}\label{E:eq35}
    \tau_{k^*}^*(\mathbf{h})=1, \quad \quad
    r_{k^*}^*(\mathbf{h})=R_{m^*}(\mathbf{h})/w_{k^*};
\end{equation}
and $\tau_k^*(\mathbf{h})=0$ and $r_k^*(\mathbf{h})=0$, $\forall k
\neq k^*$, $1 \leq k \leq K$.

\item If $\exists m^* \in \{1,2,\ldots,K_0\}$ so that $\lambda^* =
s_{m^*}(\mathbf{h})$, then
$\tilde{\tau}_{m^*}^*(\mathbf{h})=\tau_0^*$,
$\tilde{\tau}_{m^*-1}^*(\mathbf{h})=1-\tau_0^*$, and
$\tilde{\tau}_m^*(\mathbf{h})=0$, $\forall m \neq m^*, m^*-1$, $1
\leq m \leq K_0$. This implies that if $(R_{m^*}(\mathbf{h}),
C_{m^*}(\mathbf{h}))$ and $(R_{m^*-1}(\mathbf{h}),
C_{m^*-1}(\mathbf{h}))$ belongs to users $i^*$ and $j^*$,
respectively, then
\begin{equation}\label{E:eq36}
    \begin{cases}
    \tau_{i^*}^*(\mathbf{h})=\tau_0^*, &
    r_{i^*}^*(\mathbf{h})=R_{m^*}(\mathbf{h})/w_{i^*}, \\
    \tau_{j^*}^*(\mathbf{h})=1-\tau_0^*, &
    r_{j^*}^*(\mathbf{h})=R_{m^*-1}(\mathbf{h})/w_{j^*};
    \end{cases}
\end{equation}
and $\tau_k^*(\mathbf{h})=0$ and $r_k^*(\mathbf{h})=0$, $\forall k
\neq i^*,j^*$, $1 \leq k \leq K$. Note that if $m^*=1$, we let
user $i^*$ transmit with $\tau_0^*$ fraction of time and leave the
channel idle for the remaining $1-\tau_0^*$ fraction of time. If
$i^*=j^*$, then we let the same user transmit with two AMC modes,
one for $\tau_0^*$ fraction and the other for $1-\tau_0^*$
fraction of time.
\end{enumerate}

In eqs. (\ref{E:eq35}) and (\ref{E:eq36}), $\tau_0^*$ should
satisfy the average rate constraint
\begin{equation}\label{E:achieve}
    E_{\mathbf{h}} \left[\sum_{k=1}^K w_k \tau_k^*(\mathbf{h})
        r_k^*(\mathbf{h})\right]= \bar{R}.
\end{equation}}
\end{theorem}

\emph{Proof:} See Appendix D. \hfill $\Box$

The results of Theorem 3 are analogous in form with those in
\cite[Theorem 3]{Li&Goldsmith2001b}, which is a generalization of
\cite[Lemma 3]{Caire1999}. But note that the latter deal with the
delay-limited/outage capacity; while our results are for
energy-efficient resource allocation using finite-AMC-modes, a
subject not considered in \cite{Li&Goldsmith2001b, Caire1999}. In
Theorem 3, $\lambda^*$ is the water-filling level. For energy
efficiency, we should let the first derivatives
$J^{(1)}(R(\mathbf{h}))=\lambda^*$. However, since
$J(R(\mathbf{h}))$ entails a finite number of slopes, equality can
not be always achieved. Thus, our strategy is to select the
largest $J^{(1)}(R(\mathbf{h})) \leq \lambda^*$. When the largest
$J^{(1)}(R(\mathbf{h}))=s_{m^*}(\mathbf{h}) < \lambda^*$, i.e.,
$s_{m^*}(\mathbf{h}) < \lambda^* < s_{m^*+1}(\mathbf{h})$, since
the user(s) transmit more efficiently than the required power
level, we should allow transmission(s) with peak rate given by
$R_{m^*}(\mathbf{h})/w_{k^*}$. When $\lambda^* =
s_{m^*}(\mathbf{h})$, users transmit as efficiently as required;
thus arbitrary time division suffices. When $\lambda^* <
s_1(\mathbf{h})$, no transmission can be carried out as
efficiently as required, and all users defer during this fading
state. As in the infinite-codebook case, we should allow at most
two users to transmit per time block. However, here we also allow
one user to transmit in a time-sharing fashion with two AMC modes
during some time blocks.

\subsection{Discrete-Time Allocation}\label{S:dtl}

So far we have derived energy-efficient resource allocation
strategies under the assumption that the time fraction $\tau_k$
assigned to each user can be any real number in [0,1]. In
practical TDMA systems, time is usually divided with granularity
of one time unit (slot), which is determined by the available
bandwidth. Therefore, the transmission time assigned to each user
per time block has to be an integer multiple of a ``slot''. Take
the well-known GSM system as an example~\cite[Chapter 4]{FWC}. In
each narrowband channel of 200 KHz, time is divided into slots of
length 577 $\mu$s and each uplink frame is shared by the users in
a time-division manner, and consists of 8 slots. Upon regarding an
uplink frame as a frequency flat-fading time block, we can thus
only assign each user a time fraction which is an integer multiple
of 1/8. However, this extra discrete-time allocation constraint
does not affect the derived energy-efficient allocation policies.
In our policies, most of the time we should assign the entire time
block to a single user. When we occasionally allow two users to
transmit (or allow one user to transmit with two AMC modes), the
time division between them can be arbitrary. Therefore, our
energy-efficient policies can be easily adopted by practical
systems which only allow discrete-time allocation among users.

\section{Average Individual-Rate Constraints}\label{S:ind}

In resource allocation for multi-access channels, other than a
weighted average sum-rate constraint, a more general setting is
when each user has an individual average rate requirement. We next
consider power minimization under such individual rate
constraints. Without the rate-reward weight vector $\mathbf{w}$,
we let ${\cal F}'$ denote the set of all possible rate and time
allocation policies satisfying the individual rate constraints
$\left\{E_\mathbf{h}[\tau_k(\mathbf{h}) r_k(\mathbf{h})] \geq
\bar{R}_k\right\}_{k=1}^K$ and $\sum_{k=1}^K \tau_k(\mathbf{h})
=1$, $\forall \mathbf{h}$. Upon defining $\mathbf{\bar{r}} :=
[\bar{R}_1,\ldots,\bar{R}_K]^T$, the power region under the
individual rate constraints is [c.f. (\ref{E:power})]
\begin{equation}
    \tilde{{\cal P}}(\mathbf{\bar{r}}) = \bigcup_{(\mathbf{r}(\cdot),\boldsymbol{\tau}(\cdot)) \in {\cal
    F}'} {\cal P}_{TD} (\mathbf{r}(\cdot),\boldsymbol{\tau}(\cdot)),
\end{equation}
where ${\cal P}_{TD} (\mathbf{r}(\cdot),\boldsymbol{\tau}(\cdot))$
is defined as in (\ref{E:PTD}). Again, if the block length is
sufficiently large and the users are allowed to use an infinite
number of codebooks, it is easy to show that every point in
$\tilde{{\cal P}}(\mathbf{\bar{r}})$ is feasible and the region is
convex. With power cost weights $\boldsymbol{\mu}:=[\mu_1, \ldots,
\mu_K]^T$, the energy-efficient resource allocation policies solve
the optimization problem
\begin{equation}\label{E:ind1}
    \min_{\mathbf{\bar{p}}} \boldsymbol{\mu}^T \mathbf{\bar{p}}, \quad \quad
    \mbox{subject to} \quad  \mathbf{\bar{p}} \in \tilde{{\cal P}}(\mathbf{\bar{r}}).
\end{equation}
The solution $\mathbf{\bar{p}}$ yielding the optimal rate and time
allocation is on the boundary surface of $\tilde{{\cal
P}}(\mathbf{\bar{r}})$ due to its convexity. By solving
(\ref{E:ind1}) for all $\boldsymbol{\mu} \geq \mathbf{0}$, we
determine all the boundary points, and thus the whole power region
$\tilde{{\cal P}}(\mathbf{\bar{r}})$. Again, we will explicitly
characterize the optimal resource allocation policies and the
resultant boundary point for $\boldsymbol{\mu}> \mathbf{0}$. By
letting some of the power cost weights approach 0, we can get
arbitrarily close to the extreme points.

\subsection{Infinite-Codebooks}

When the user can utilize an infinite number of codebooks to
achieve channel capacity per fading state, the optimization in
(\ref{E:ind1}) is equivalent to
\begin{equation}\label{E:ind2}
    \begin{cases}
        \min_{\mathbf{r}(\cdot),\boldsymbol{\tau}(\cdot)} E_{\mathbf{h}} \left[ \sum_{k=1}^K \mu_k
        \frac{\tau_k(\mathbf{h})}{h_k} (2^{r_k(\mathbf{h})}-1)\right]
        \\
        \mbox{subject to} \quad E_{\mathbf{h}} \left[ \tau_k(\mathbf{h})
        r_k(\mathbf{h})\right] \geq \bar{R}_k, \;\; k=1,\ldots, K, \\
        \hspace{2cm}  \sum_{k=1}^K \tau_k(\mathbf{h}) =1, \;\; \forall
        \mathbf{h}.
    \end{cases}
\end{equation}
The counterpart of Theorem 2 in this case is given by:

\begin{theorem}
{\it For any $\boldsymbol{\mu} > \mathbf{0}$, there exists a
$\boldsymbol{\lambda}^*:= [\lambda_1^*, \ldots, \lambda_K^*]^T >
\mathbf{0}$, and optimal rate and time allocation policies
$\mathbf{r}^*(\cdot)$ and $\boldsymbol{\tau}^*(\cdot)$ in
(\ref{E:ind2}), such that for each $\mathbf{h}$,
$\mathbf{r}^*(\mathbf{h})$ and $\boldsymbol{\tau}^*(\mathbf{h})$
solve
\begin{equation}\label{E:ind3}
    \begin{cases}
        \min_{\mathbf{r}(\mathbf{h}), \boldsymbol{\tau}(\mathbf{h})} \sum_{k=1}^K \tau_k(\mathbf{h})
        f_k(r_k(\mathbf{h})) \\
        \mbox{subject to} \quad \sum_{k=1}^K \tau_k(\mathbf{h})
        =1,
    \end{cases}
\end{equation}
where now $f_k(r_k(\mathbf{h})):=\frac{\mu_k}{h_k}
(2^{r_k(\mathbf{h})}-1) - \lambda_k^* r_k(\mathbf{h})$. Since
$f_k(r_k(\mathbf{h}))$ is convex in $r_k(\mathbf{h})$, it attains
its minimum at $r_{k,\min} (\mathbf{h}): = \left[ \log \lambda_k^*
- \log \frac{\ln 2\mu_k}{h_k} \right]_+$. Moreover, we have
$\mathbf{r}^*(\mathbf{h})$ and $\boldsymbol{\tau}^*(\mathbf{h})$
as follows.

\begin{enumerate}
\item If functions $f_k(r_{k,\min} (\mathbf{h}))$, $k=1,\ldots,
K$, have a single minimum $f_i(r_{i,\min} (\mathbf{h}))$, i.e.,
$$i=\arg \; \min_k \; f_k(r_{k,\min} (\mathbf{h})),$$ then $\forall k
\neq i$, $k \in [1, K]$, $r_k^*(\mathbf{h}) = 0$,
$\tau_k^*(\mathbf{h}) = 0$, and
\begin{equation}\label{E:eq42}
    r_i^*(\mathbf{h}) = r_{i,\min} (\mathbf{h}), \qquad \tau_i^*(\mathbf{h})=
    1.
\end{equation}

\item If functions $f_k(r_{k,\min} (\mathbf{h}))$, $k=1,\ldots,
K$, have multiple minima $\left\{f_{i_j}(r_{i_j,\min}
(\mathbf{h}))\right\}_{j=1}^J$, then
\begin{equation}\label{E:eq43}
r_{i_j}^*(\mathbf{h}) = r_{i_j, \min} (\mathbf{h}), \qquad
\tau_{i_j}^*(\mathbf{h}) = \tau_j^*,
\end{equation}
with arbitrary $\sum_{j=1}^{J} \tau_j^* =1$, and $\forall k \neq
i_j$, $k \in [1, K]$, $r_k^*(\mathbf{h}) = 0$ and
$\tau_k^*(\mathbf{h}) = 0$.
\end{enumerate}
In (\ref{E:eq42}) and (\ref{E:eq43}), $\boldsymbol{\lambda}^*$ and
$\{\tau_j^*\}_{j=1}^J$ are obtained by satisfying the individual
rate constraints
\begin{equation}\label{E:vlambda}
E_{\mathbf{h}} \left[ \tau_k^*(\mathbf{h})
r_k^*(\mathbf{h})\right] = \bar{R}_k, \quad k=1,\ldots, K.
\end{equation}}
\end{theorem}

\emph{Proof:} See Appendix E. \hfill $\Box$

If we regard $f_k(r_{k,\min} (\mathbf{h}))$ as a channel quality
indicator (the smaller the better) for user $k$, Theorem 4 asserts
that for each time block, we should only allow the user with the
``best'' channel to transmit. When there are multiple users with
``best'' channels, arbitrary time division among them suffices.
Therefore, our resource allocation strategies are ``greedy'' ones.
Note that $f_k(r_{k,\min}(\mathbf{h}))$ contains $\lambda_k^*$.
This implies that the user having smallest
$f_k(r_{k,\min}(\mathbf{h}))$ actually has the
rate-constraint-controlled ``best'' channel.

Rate maximization under individual power constraints was pursued
in~\cite{Tse&Hanly1998}, where it was shown that superposition
codes and successive decoding should be employed and that greedy
water-filling based on a polymatroid structure provides the
optimal resource allocation. In our power-efficient TDMA setting,
we do not have such a polymatroid structure. Albeit in different
forms than those in~\cite{Tse&Hanly1998}, our strategies can be
also implemented through a greedy water-filling approach. To
obtain the optimal allocation policies in Theorem 4, we need to
calculate the Lagrange multiplier vector $\boldsymbol{\lambda}^*$.
Although a $K$-dimensional search can be used to directly compute
$\boldsymbol{\lambda}^*$ from (\ref{E:vlambda}), it is
computationally inefficient when $K$ is large. Next, we show that
an iterative algorithm from~\cite{Tse&Hanly1998} can be adopted to
calculate $\boldsymbol{\lambda}^*$. Before that, by the strict
convexity of exponential functions and the fact that
non-uniqueness of the time allocation $\boldsymbol{\tau}^*(\cdot)$
occurs with probability 0 when $F(\mathbf{h})$ is continuous, we
can argue as in \cite[Lemma 3.15]{Tse&Hanly1998} to establish the
following lemma:

\begin{lemma}
{\it Given a positive power weight vector $\boldsymbol{\mu}$,
there exists a unique $\mathbf{\bar{p}}^*$ which minimizes
$\boldsymbol{\mu}^T \mathbf{\bar{p}}$, and there is a unique
Lagrange vector $\boldsymbol{\lambda}^*$ such that the optimal
rate and time allocation satisfy the average individual rate
constraints.}
\end{lemma}

To gain more insight, let us look at a special case where the
fading processes of users are independent. If $F_k(\cdot)$ stands
for the cumulative distribution function (cdf) of user $k$'s
fading channel, define $g_i(x):= \frac{\lambda_i^*}{\ln
2}-\frac{\mu_i}{x} -\lambda_i^* \log \frac{\lambda_i^*x}{\ln
2\mu_i}$, and let $s_{i,k}(z)$ denote the solution to
$g_{i,k}(x,z):=g_i(x)-g_k(z)=0$.

\begin{corollary}
{\it If the fading processes of users are independent, the optimal
solution $\mathbf{\bar{p}}^*$ to (\ref{E:ind1}) for a given
$\boldsymbol{\mu}>\mathbf{0}$ can be obtained as
\begin{equation}
    \bar{P}_k^* = \int_{\frac{\ln 2\mu_k}{\lambda_k^*}}^{\infty} \;
    \left(\frac{\lambda_k^*}{\ln 2 \mu_k} - \frac{1}{z}\right) \prod_{i \neq k} F_i (s_{i,k}(z))\;
    dF_k(z), \qquad k=1,\ldots, K;
\end{equation}
where the vector $\boldsymbol{\lambda}^*$ is the unique solution
to the equations
\begin{equation}\label{E:lambdaj}
    \int_{\frac{\ln 2\mu_k}{\lambda_k^*}}^{\infty} \;
    \log \frac{\lambda_k^*z}{\ln 2\mu_k} \prod_{i \neq k} F_i (s_{i,k}(z))\;
    dF_k(z) = \bar{R}_k, \qquad k=1,\ldots, K.
\end{equation}}
\end{corollary}

\emph{Proof:} By definition, we have
\begin{equation}
    r_{k,\min}(\mathbf{h}) =\left[\log \lambda_k^*-\log
    \frac{\ln 2\mu_k}{h_k} \right]_+ =
    \begin{cases}
        \log \frac{\lambda_k^*h_k}{\ln 2\mu_k}, &  h_k
        \geq \frac{\ln 2\mu_k}{\lambda_k^*} \\
        0, & 0 \leq h_k \leq \frac{\ln 2\mu_k}{\lambda_k^*}
    \end{cases},
\end{equation}
\begin{equation}
    f_k(r_{k,\min}(\mathbf{h})) =
    \begin{cases}
        \frac{\lambda_k^*}{\ln 2}-\frac{\mu_k}{h_k} - \lambda_k^* \log \frac{\lambda_k^*h_k}{\ln 2\mu_k}, &
        h_k
        \geq \frac{\ln 2\mu_k}{\lambda_k^*} \\
        0, & 0 \leq h_k \leq \frac{\ln 2\mu_k}{\lambda_k^*}
    \end{cases},
\end{equation}
and
\begin{equation}\label{E:partial}
    \frac{\partial f_k(r_{k,\min}(\mathbf{h}))}{\partial h_k} =
    \begin{cases}
        \frac{1}{h_k} \left( \frac{\mu_k}{h_k}- \frac{\lambda_k^*}{\ln 2} \right), &
        h_k
        \geq \frac{\ln 2\mu_k}{\lambda_k^*} \\
        0, & 0 \leq h_k \leq \frac{\ln 2\mu_k}{\lambda_k^*}
    \end{cases}.
\end{equation}
Since $\frac{\partial f_k(r_{k,\min}(\mathbf{h}))}{\partial h_k}
\leq 0$, we have from (\ref{E:partial}) [c.f. Theorem 4]
\begin{eqnarray}
    \nonumber
    E_{\mathbf{h}} \left[ \tau_k^*(\mathbf{h})r_k^*(\mathbf{h})\right]
    & = & E_{\mathbf{h}} \left[ r_k^*(\mathbf{h}) \mathbf{I}_{\{ f_i(r_{i,\min}(\mathbf{h}))> f_k(r_{k,\min}(\mathbf{h})),
    \;\; \forall i \}} \right] \\
    & = &  \int_{\frac{\ln 2\mu_k}{\lambda_k^*}}^{\infty} \;
    \log \frac{\lambda_k^*z}{\ln 2\mu_k} \prod_{i \neq k} F_i (s_{i,k}(z))\;
    dF_k(z), \label{E:ER}
\end{eqnarray}
and
\begin{eqnarray}
    \nonumber
    \bar{P}_k^* & = &  E_{\mathbf{h}} \left[ \frac{1}{h_k} (2^{r_{k,\min}(\mathbf{h})}-1)
    \mathbf{I}_{\{ f_i(r_{i,\min}(\mathbf{h}))> f_k(r_k^m(\mathbf{h})),
    \;\; \forall i \}} \right] \\
    & = & \int_{\frac{\ln 2\mu_k}{\lambda_k^*}}^{\infty} \;
    \left(\frac{\lambda_k^*}{\ln 2 \mu_k}- \frac{1}{z}\right) \prod_{i \neq k} F_i (s_{i,k}(z))\;
    dF_k(z). \label{E:EP}
\end{eqnarray}
The corollary thus follows from (\ref{E:ER}) and (\ref{E:EP}).
\hfill $\Box$

What left to obtain the optimum $\{\bar{P}_k^*\}_{k=1}^K$, is to
specify $\{\lambda_k^*\}_{k=1}^K$. We accomplish this by modifying
the corresponding iterative algorithm in~\cite{Tse&Hanly1998}.

\begin{algorithm}
Let $\boldsymbol{\lambda}(0)> \mathbf{0}$ be an arbitrary initial
rate-reward vector. Given the $l$th iterate
$\boldsymbol{\lambda}(l)$, the $(l+1)$st iterate
$\boldsymbol{\lambda}(l+1)$ is calculated as follows. For each $k
\in [1, K]$, $\lambda_k(l+1)$ is the unique rate-reward weight for
the $k$th user such that the average rate of user $k$ is
$\bar{R}_k$ under the optimal rate and time allocation policies,
when the rate-reward weights of other users remain fixed at
$\boldsymbol{\lambda}(l)$.
\end{algorithm}

In the independent fading case, $\lambda_k(l+1)$ is the unique
root $\lambda_k^*$ of (\ref{E:lambdaj}), which can be numerically
solved if the fading statistics are known. Let
$\mathbf{\bar{r}}(\boldsymbol{\lambda}(l)):=[\bar{R}_1(\boldsymbol{\lambda}(l)),\ldots,
\bar{R}_K(\boldsymbol{\lambda}(l))]^T$ and
$\mathbf{\bar{p}}(\boldsymbol{\lambda}(l)):=[\bar{P}_1(\boldsymbol{\lambda}(l)),\ldots,
\bar{P}_K(\boldsymbol{\lambda}(l))]^T$ denote the rates and powers
of users given by Theorem 4, respectively, when the Lagrange
multiplier is $\boldsymbol{\lambda}(l)$. We can then prove that:

\begin{theorem}
{\it Given the average rate constraint $\mathbf{\bar{r}}$, if
$\mathbf{\bar{p}}^*$ is the optimal power vector corresponding to
the cost vector $\boldsymbol{\mu}$, and $\boldsymbol{\lambda}^*$
is the rate-reward vector satisfying (\ref{E:vlambda}), then
\begin{equation}
    \boldsymbol{\lambda}(l) \rightarrow \boldsymbol{\lambda}^*, \qquad \mbox{as} \;\; l
    \uparrow \infty;
\end{equation}
and hence $\mathbf{\bar{p}}(\boldsymbol{\lambda}(l)) \rightarrow
\mathbf{\bar{p}}^*$ and $\mathbf{\bar{r}}(\boldsymbol{\lambda}(l))
\rightarrow \mathbf{\bar{r}}$.}
\end{theorem}

\emph{Proof:} See Appendix F. \hfill $\Box$

Algorithm 4 is also widely used for power control in IS-95 CDMA
systems and its convergence is guaranteed~\cite[Chapter 4]{FWC}.
For the power minimization in our TDMA setting, Theorem 5 ensures
convergence of Algorithm 4 in finding the optimal Lagrange
multiplier $\boldsymbol{\lambda}^*$. The proof is analogous to
that of \cite[Theorem 4.3]{Tse&Hanly1998}, except for some
necessary modifications. With Theorem 4 and Algorithm 4, we
determine the energy-efficient rate and time allocation policies
under individual rate constraints. Our policies are greedy in the
sense that most of the time we allow a single user with the
``best'' channel to transmit, and occasionally we assign time to
multiple users with ``equally best'' channels at a given time
block. Note that water-filling may result in no user transmissions
for some fading states, where all channels are in deep fading.

\subsection{Finite AMC Modes}

Next, we investigate optimal resource allocation under individual
average rate constraints for the case when each user can only
support a finite number of AMC modes. As in Sec.\ \ref{S:AMC}, for
user $k \in [1, K]$, an AMC mode corresponds to a rate-power pair
$\{(\rho_{k,l}, p_{k,l})\}_{l=1}^{M_k}$, where $M_k$ is the number
of AMC modes. By time-sharing, a user can support continuous rates
up to a maximum value determined by the highest-rate AMC mode
$\rho_{k,M_k}$. By defining $\Upsilon_k (x)$ as in
(\ref{E:upsilon}),
the new power region is given by
\begin{equation}
    \tilde{{\cal P}}'(\mathbf{\bar{r}}) = \bigcup_{(\mathbf{r}(\cdot),\boldsymbol{\tau}(\cdot)) \in {\cal
    F}'} {\cal P}_{TD}' (\mathbf{r}(\cdot),\boldsymbol{\tau}(\cdot)),
\end{equation}
where ${\cal F}'$ denotes the set of all possible rate and time
allocation policies satisfying the individual rate constraints,
and ${\cal P}_{TD}' (\mathbf{r}(\cdot),\boldsymbol{\tau}(\cdot))$
is defined as in (\ref{E:wsum2}). It is easy to show that the
region ${\cal P}'(\mathbf{\bar{r}})$ is feasible and convex. The
optimization problem thus becomes
\begin{equation}\label{E:optamc}
    \min_{\mathbf{\bar{p}}} \boldsymbol{\mu}^T \mathbf{\bar{p}}, \quad \quad
    \mbox{subject to} \quad  \mathbf{\bar{p}} \in \tilde{{\cal P}}'(\mathbf{\bar{r}}).
\end{equation}
Using $\Upsilon_k (x)$, problem (\ref{E:optamc}) is equivalent to
\begin{equation}\label{E:optamc2}
    \begin{cases}
        \min_{\mathbf{r}(\cdot),\boldsymbol{\tau}(\cdot)} E_{\mathbf{h}} \left[ \sum_{k=1}^K \mu_k \tau_k(\mathbf{h})
        \Upsilon_k (r_k(\mathbf{h}))\right]
        \\
        \mbox{subject to} \quad E_{\mathbf{h}} \left[ \tau_k(\mathbf{h})
        r_k(\mathbf{h})\right] \geq \bar{R}_k, \;\; k=1,\ldots, K, \\
        \hspace{2cm}  \sum_{k=1}^K \tau_k(\mathbf{h}) =1, \;\; \forall
        \mathbf{h}.
    \end{cases}
\end{equation}
Since every point of $\Upsilon_k (r_k(\mathbf{h}))$ can be
achieved by time-sharing between points $(\rho_{k,l},
p_{k,l}/h_k)$, finding the optimal resource allocation strategies
for (\ref{E:optamc2}) is equivalent to solving
\begin{equation}\label{E:optamc3}
    \begin{cases}
        \min_{\boldsymbol{\tilde{\tau}}(\mathbf{h})} \; \sum_{k=1}^{K} \; E_{\mathbf{h}}
        \left[\sum_{l=0}^{M_k} \mu_k
        \frac{\tilde{\tau}_{k,l}(\mathbf{h})}{h_k} p_{k,l} \right] \\
        \mbox{subject to} \quad E_{\mathbf{h}}
        \left[\sum_{l=0}^{M_k} \tilde{\tau}_{k,l}(\mathbf{h})
        \rho_{k,l}\right] \geq \bar{R}_k, \;\; k=1,\ldots, K, \\
        \hspace{2cm}  \sum_{k=1}^{K} \; \sum_{l=0}^{M_k} \tilde{\tau}_{k,l}(\mathbf{h}) = 1, \;\; \forall
        \mathbf{h}.
    \end{cases}
\end{equation}
The counterpart of Theorem 3 under individual rate constraints is
now:

\begin{theorem}
{\it If $\mathbf{\bar{r}}$ is feasible $\forall \mathbf{h}$, we
have the optimal solution $\tilde{\tau}_{k,l}^*(\mathbf{h})$ ($k
\in [1, K]$, $l \in [0, M_k]$) to (\ref{E:optamc3}), and
subsequently the optimal allocation $r_k^*(\mathbf{h})$ and
$\tau_k^*(\mathbf{h})$ for (\ref{E:optamc2}) as follows. Given a
positive $\boldsymbol{\lambda}^*:= [\lambda_1^*, \ldots,
\lambda_K^*]^T$, for each fading state $\mathbf{h}$, let
$l_k^*:=\max \; \{l:\; \mu_k \gamma_{k,l}/h_k \leq \lambda_k^*\}$
($l_k^*=0$ if no such $l$) and $C_{k,l}:=\mu_k p_{k,l}/h_k$, and
define $\varphi_k(\mathbf{h}):=C_{k,l_k^*}-\lambda_k^*
\rho_{k,l_k^*}$.
\begin{enumerate}
\item If $\{\varphi_k(\mathbf{h})\}_{k=1}^K$ have a single minimum
$\varphi_i(\mathbf{h})$, i.e., $i=\arg \; \min_k \;
\varphi_k(\mathbf{h})$, then $\tilde{\tau}_{i,l_i^*}=1$ and all
other $\tilde{\tau}_{k,l}=0$. Consequently,
\begin{equation}\label{E:eq57}
    r_i^*(\mathbf{h})=\rho_{i,l_i^*}, \qquad
    \tau_i^*(\mathbf{h})=1;
\end{equation}
and $\forall k \neq i$, $k \in [1, K]$, $r_k^*(\mathbf{h})=0$ and
$\tau_k^*(\mathbf{h})=0$.

\item If $\{\varphi_k(\mathbf{h})\}_{k=1}^K$ have multiple minima
$\left\{\varphi_{i_j}(\mathbf{h})\right\}_{j=1}^J$, then
$\tilde{\tau}_{i_j,l_{i_j}^*}=\tau_j^*$ with arbitrary
$\sum_{j=1}^{J} \tau_j^* =1$, and all other
$\tilde{\tau}_{k,l}=0$. Consequently,
\begin{equation}\label{E:eq58}
    r_{i_j}^*(\mathbf{h})=\rho_{i_j,l_{i_j}^*}, \qquad
    \tau_{i_j}^*(\mathbf{h})=\tau_j^*,
\end{equation}
and $\forall k \neq i_j$, $k \in [1, K]$, $r_k^*(\mathbf{h}) = 0$
and $\tau_k^*(\mathbf{h}) = 0$.
\end{enumerate}
In (\ref{E:eq57}) and (\ref{E:eq58}), $\boldsymbol{\lambda}^*$ and
$\{\tau_j^*\}_{j=1}^J$ should satisfy the individual rate
constraints
\begin{equation}\label{E:vlambda2}
E_{\mathbf{h}} \left[ \tau_k^*(\mathbf{h})
r_k^*(\mathbf{h})\right] = \bar{R}_k, \quad k=1,\ldots, K.
\end{equation}
Moreover, $\boldsymbol{\lambda}^*$ is almost surely unique and can
be iteratively computed by Algorithm 4.}
\end{theorem}

\emph{Proof:} See Appendix G. \hfill $\Box$

Theorem 6 shows that our policies with finite number of AMC-modes
are still greedy ones. For user $k$ at fading state $\mathbf{h}$,
the C$\thicksim$R function is given by
$f_k(r_k(\mathbf{h})):=\mu_k \Upsilon_k (r_k(\mathbf{h})) -
\lambda_k^* r_k(\mathbf{h})$. It is clear that this function
attains its minimum at $\rho_{k,l_k^*}$. Then comparing the
channel quality indicators $\varphi_k(\mathbf{h})$ for all the
users, we determine which users have the ``best'' channel and
assign resources accordingly. Note that when $l_k^*=0$, the user
should remain silent. If at a fading state, this user happens to
have the ``best" channel, we should let all users defer at this
block. When $\mu_k \gamma_{k,l_k^*}/h_k = \lambda_k^*$, the whole
line between $(\rho_{k,l_k^*-1}, p_{k,l_k^*-1}/h_k)$ and
$(\rho_{k,l_k^*}, p_{k,l_k^*}/h_k)$ in $\Upsilon_k
(r_k(\mathbf{h}))$ achieves the minimum of $f_k(r_k(\mathbf{h}))$.
Although in Theorem 6 we let user $k$ transmit (if permitted by
the policies) with rate $\rho_{k,l_k^*}$, the complete solutions
should allow this user to transmit with arbitrary time-sharing
between $\rho_{k,l_k^*-1}$ and $\rho_{k,l_k^*}$, as the
optimization under a weighted sum-rate constraint in Theorem 3.
Summarizing, our greedy polices may result in no transmissions,
user(s) transmitting with one AMC mode, or user(s) transmitting in
a time-sharing fashion with two AMC modes per fading block.

For the special case where the fading processes of users are
independent, let $s_{i,k}(z)$ denote the solution to
$g_{i,k}'(x,z):=\varphi_i(x)-\varphi_k(z)=0$. Note that
$\varphi_i(x)$ is also a function of $\lambda_i^*$. Using Theorem
6 and mimicking the proof of Corollary 1, we can establish the
following corollary.

\begin{corollary}
{\it If the fading processes of users are independent, the optimal
solution $\mathbf{\bar{p}}^*$ to (\ref{E:optamc}) for a given
$\boldsymbol{\mu} > \mathbf{0}$ can be obtained as
\begin{equation}
    \bar{P}_k^* = \int_{\frac{\mu_k p_{k,1}}{\lambda_k^* \rho_{k,1}}}^{\infty} \;
    (p_{k,l_k^*}/z) \prod_{i \neq k} F_i (s_{i,k}(z))\;
    dF_k(z), \qquad k=1,\ldots, K;
\end{equation}
where the vector $\boldsymbol{\lambda}^*$ is the unique solution
to the equations
\begin{equation}\label{E:lambdaj2}
    \int_{\frac{\mu_k p_{k,1}}{\lambda_k^* \rho_{k,1}}}^{\infty} \;
    \rho_{k,l_k^*} \prod_{i \neq k} F_i (s_{i,k}(z))\;
    dF_k(z) = \bar{R}_k, \qquad k=1,\ldots, K.
\end{equation}}
\end{corollary}

Some comments are now in order: 1) in the finite-AMC-mode case,
the rate $\rho_{k,l}$ of user $k$ is maintained with a prescribed
SEP, and $p_{k,l}$ is the corresponding minimum received power
under the SEP constraint; 2) our policies for the finite-AMC-mode
case could only require {\it quantized} CSI at the transmitters;
and 3) following the arguments of Sec.\ \ref{S:dtl}, it is clear
that the derived policies for infinite codebooks and finite number
of AMC-modes apply to systems where only discrete-time allocations
are allowed among users.

\section{Frequency Selective Channels}\label{S:selective}

The optimal resource allocation policies in the previous sections
are derived for frequency-flat block fading channels encountered
with narrow-band communications. In this section we extend our
results to frequency-selective fading channels, which are often
encountered in wide-band communication systems.

Supposing that the channel varies very slowly relative to the
multipath delay spread, it can be decomposed into a set of
parallel time-invariant Gaussian multi-access channels in the
spectral domain~\cite{Cheng&Verdu1993}. We consider an $K$-user
spectral Gaussian block fading TDMA channel with continuous fading
spectra $H_1(f,\boldsymbol{\omega})$,
$H_2(f,\boldsymbol{\omega})$, $\ldots$,
$H_K(f,\boldsymbol{\omega})$, where frequency $f$ ranges over the
system bandwidth and $\boldsymbol{\omega}$ is the fading state at
a given time block. Let $r_k(f,\boldsymbol{\omega})$ and
$\tau_k(f,\boldsymbol{\omega})$ denote the rate and fraction of
time allocated to user $k$ at frequency $f$ and fading state
$\boldsymbol{\omega}$. For the weighted sum-rate constraint
optimization, the constraint is now given by
\begin{equation}\label{E:freq}
    \int_{\boldsymbol{\omega}}\left[\sum_{k=1}^K \int_{f} w_k
    \tau_k(f,\boldsymbol{\omega})
    r_k(f,\boldsymbol{\omega})\; df \right]\; dF(\boldsymbol{\omega}) \geq \bar{R},
    \qquad \sum_{k=1}^K \tau_k(f,\boldsymbol{\omega}) =1, \;\; \forall f,
    \boldsymbol{\omega}.
\end{equation}
Let the set ${\cal F}_{\mathbf{w}}$ consist of all possible rate
and time allocation policies satisfying (\ref{E:freq}), and take
the infinite-codebook case for illustration. The power region for
this TDMA channel is given by
\begin{equation}
    \bigcup_{(\mathbf{r}(\cdot),\boldsymbol{\tau}(\cdot)) \in {\cal F}_{\mathbf{w}}} \; \left\{ \mathbf{\bar{p}}:
    \bar{P}_k
    \geq \int_{\boldsymbol{\omega}} \left[ \int_f \frac{\tau_k(f,\boldsymbol{\omega})}{h_k(f,\boldsymbol{\omega})}
    \left(2^{r_k(f,\boldsymbol{\omega})}-1 \right) \;df \right] dF(\boldsymbol{\omega}) , \quad 1 \leq k \leq
    K  \right\}.
\end{equation}
For a finite number of AMC-modes and the individual rate
constraint optimization, we can similarly define the corresponding
power regions. Subsequently, the optimal resource allocation
strategies can be obtained from the previous results by replacing
the fading state $\mathbf{h}$ with the frequency and fading state
pair $(f,\boldsymbol{\omega})$ to determine power regions for
frequency-selective channels. That is, we should employ the
previous allocation policies for each $(f,\boldsymbol{\omega})$,
and then implement water-filling across both frequency $f$ and
fading state $\boldsymbol{\omega}$ realizations to determine
$\lambda^*$ (or vector $\boldsymbol{\lambda}^*$).

\section{Numerical Results}\label{S:simul}

In this section, we present numerical results of our
energy-efficient resource allocation for a two-user Rayleigh
flat-fading TDMA channel. The available system bandwidth is
$B=100$ KHz, and the AWGN has two-sided power spectral density
$N_0$ Watts/Hz. The user fading processes are independent and the
state $h_k$, $k=1,2$, is subject to Rayleigh fading with mean
$\bar{h}_k$. Clearly, the signal-to-noise ratio (SNR) for user $k$
is defined as $\bar{h}_k/(N_0B)$.

\begin{figure}
\centering \epsfig{file=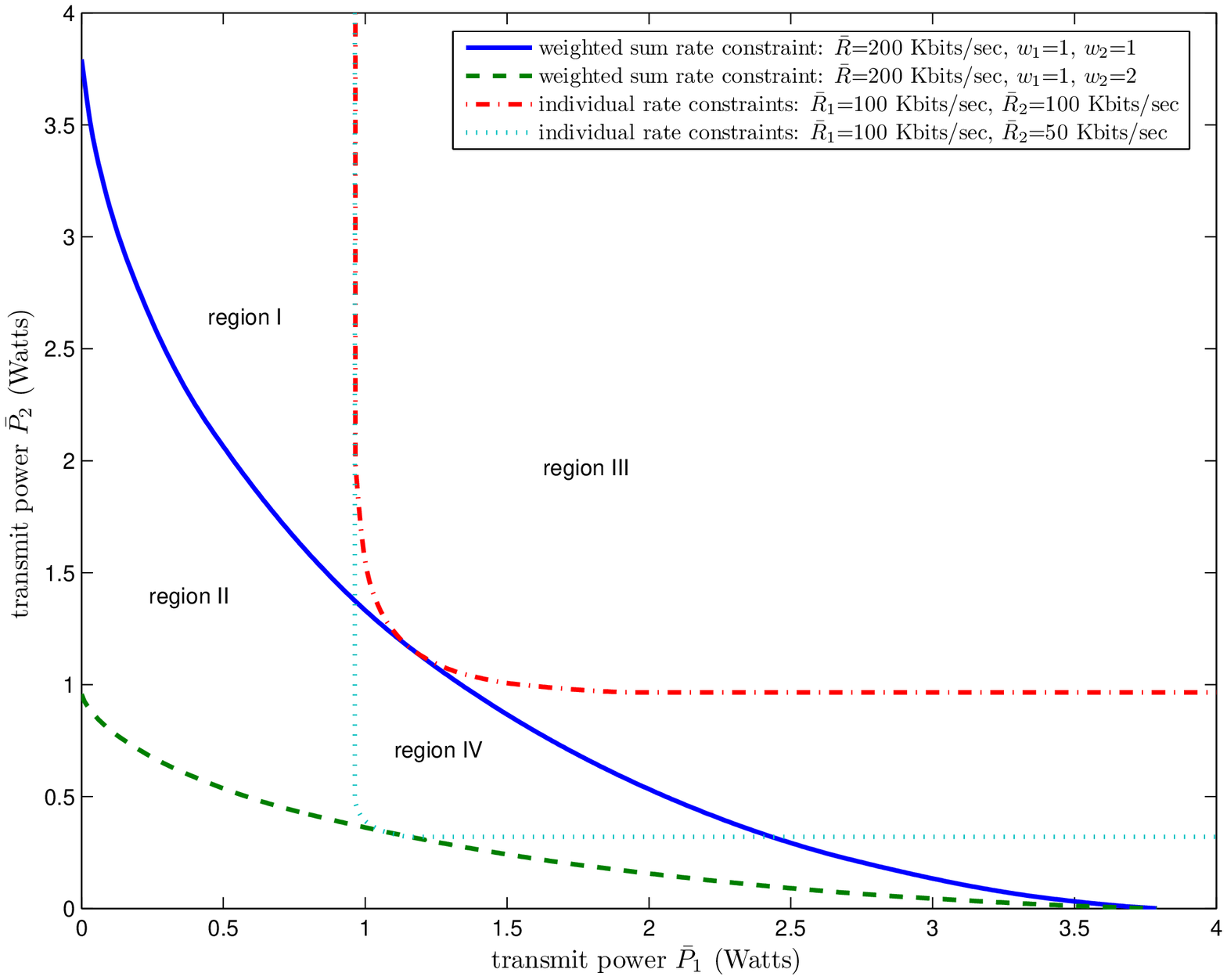, width=0.7\textwidth} \caption{
Power regions for the infinite-codebook case when two users have
identical SNRs: $\bar{h}_1/(N_0B)=\bar{h}_2/(N_0B)=0$ dBW. }
\label{F:pr1}
\end{figure}

Supposing $\bar{h}_k/(N_0B)=0$ dBW, $k=1,2$, we test
energy-efficient resource allocation under a weighted sum average
rate constraint $\bar{R}=200$ Kbits/sec, for two different sets of
rate-reward weights: i) $w_1=1$, $w_2=1$, and ii) $w_1=1$,
$w_2=2$; and the resource allocation under two different sets of
individual rate constraints: iii) $\bar{R}_1=100$ Kbits/sec,
$\bar{R}_2=100$ Kbits/sec, and iv) $\bar{R}_1=100$ Kbits/sec,
$\bar{R}_2=50$ Kbits/sec. Fig.\ \ref{F:pr1} depicts the power
regions of the Rayleigh fading TDMA fading channels for the
infinite-codebook case. It is seen that power regions I and III
under the weighted sum rate constraint i) and under individual
rate constraints iii) are symmetric with respect to the line
$\bar{P}_2=\bar{P}_1$. Since the individual rate constraints can
be seen as a realization of the weighted sum-rate constraint,
i.e., $w_1\bar{R}_1+w_2\bar{R}_2=\bar{R}$, the power region I
contains power region III. It is clear that when $\mu_1=\mu_2$,
due to the symmetry in channel quality and rate-reward weights
between the two users, the optimal resource allocation should
result in $\bar{R}_1=\bar{R}_2$ under the weighted sum rate
constraint. For this reason, the two power regions touch each
other in this case. The relation between power regions II and IV
under the weighted sum average rate constraint ii) and under
individual rate constraints iv), are similar. They are not
symmetric with respect to $\bar{P}_2=\bar{P}_1$ due to the
unbalanced rate-reward weights or individual rate constraints.
Power region II contains power region IV, and the two regions
touch each other at one point.

\begin{figure}
\centering \epsfig{file=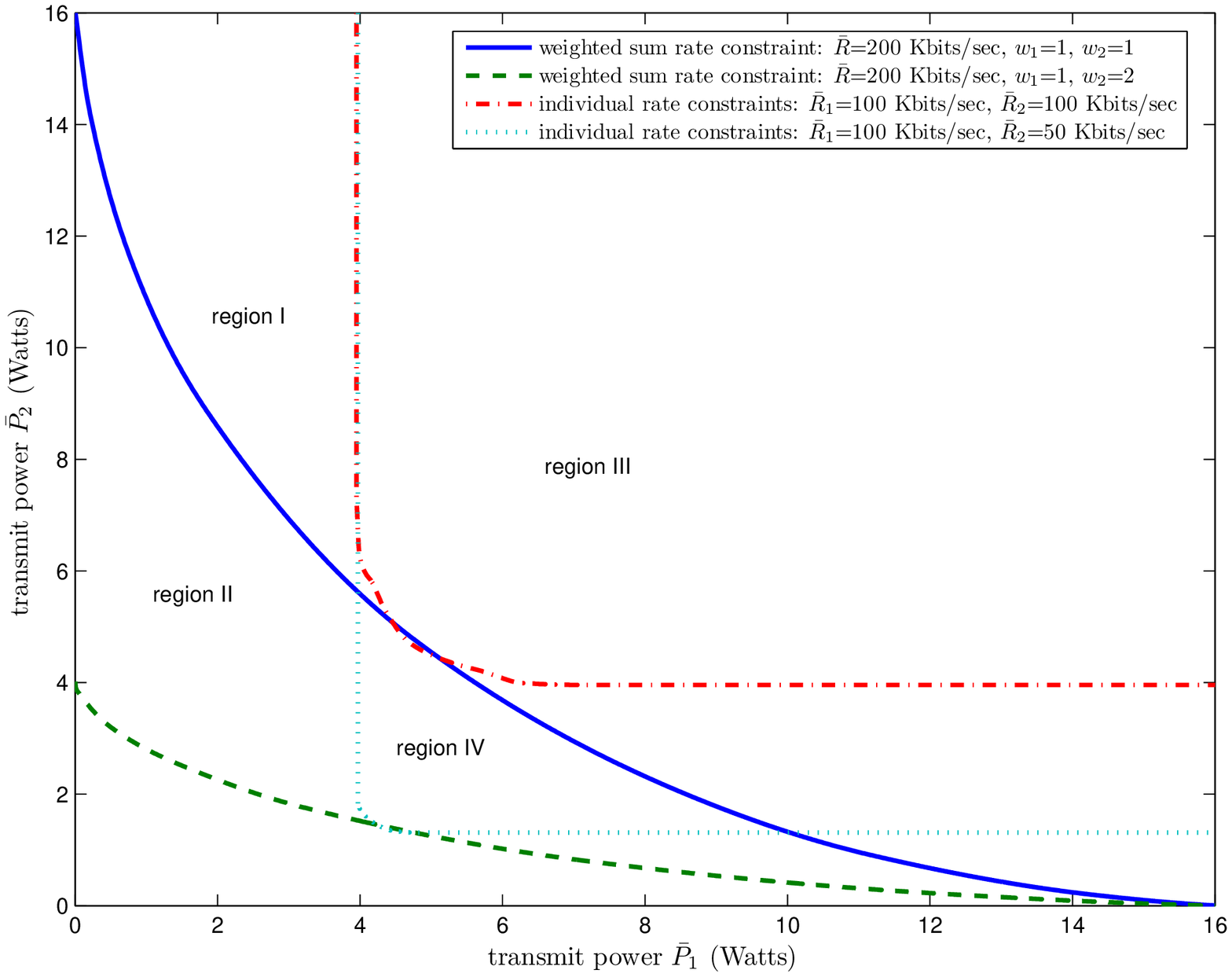, width=0.7\textwidth} \caption{
Power regions for the finite-AMC-mode case when two users have
identical SNRs: $\bar{h}_1/(N_0B)=\bar{h}_2/(N_0B)=0$ dBW. }
\label{F:pr2}
\end{figure}

For the finite-AMC-mode case, we assume henceforth that each user
supports three $M$-ary quadrature amplitude modulation (QAM)
modes: 4-QAM, 16-QAM and 64-QAM. For these rectangular signal
constellations, the SEP is given by~\cite[Chapter 5]{dcomm}
\begin{equation}\label{E:PM}
    P_M = 1-(1-P_{\sqrt{M}})^2,
\end{equation}
where
    $P_{\sqrt{M}} = 2\left( 1-\frac{1}{\sqrt{M}}\right)
    Q\left(\sqrt{\frac{3}{M-1}\frac{p_k h_k}{N_0B}}\right),$
and $Q(x):=\int_{x}^{\infty} (1/\sqrt{2\pi})e^{-y^2/2} dy$ is the
Marcum's Q-function. From (\ref{E:PM}), we determine the
rate-power pair $\{(\rho_{k,l}, p_{k,l})\}_{l=1}^3$ for user
$k=1,2$. The corresponding power regions I-IV under the
constraints i)-iv) for this finite-AMC-mode case with prescribed
SEP = $10^{-3}$ are shown in Fig.\ \ref{F:pr2}. Similar trends as
in Fig.\ \ref{F:pr1} are observed. However, the power regions
shrink since more power is required to achieve the same
transmission rate in the finite-AMC-mode case relative to that in
the infinite-codebook case.

\begin{figure}
\centering \epsfig{file=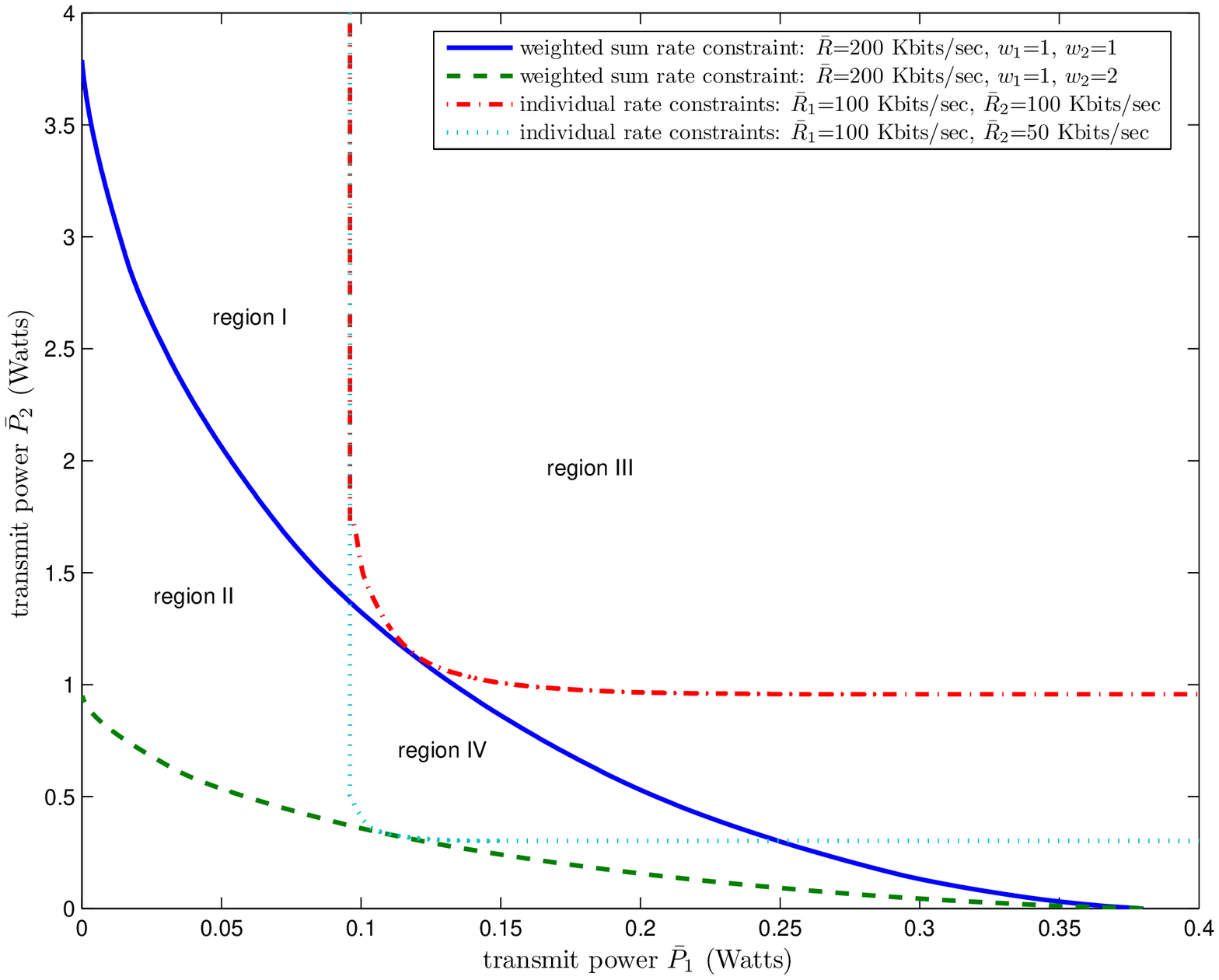, width=0.7\textwidth} \caption{
Power regions for the infinite-codebook case when two users have
10 dB difference in SNRs: $\bar{h}_1/(N_0B)=10$ dBW, and
$\bar{h}_2/(N_0B)=0$ dBW. } \label{F:pr3}
\end{figure}

\begin{figure}
\centering \epsfig{file=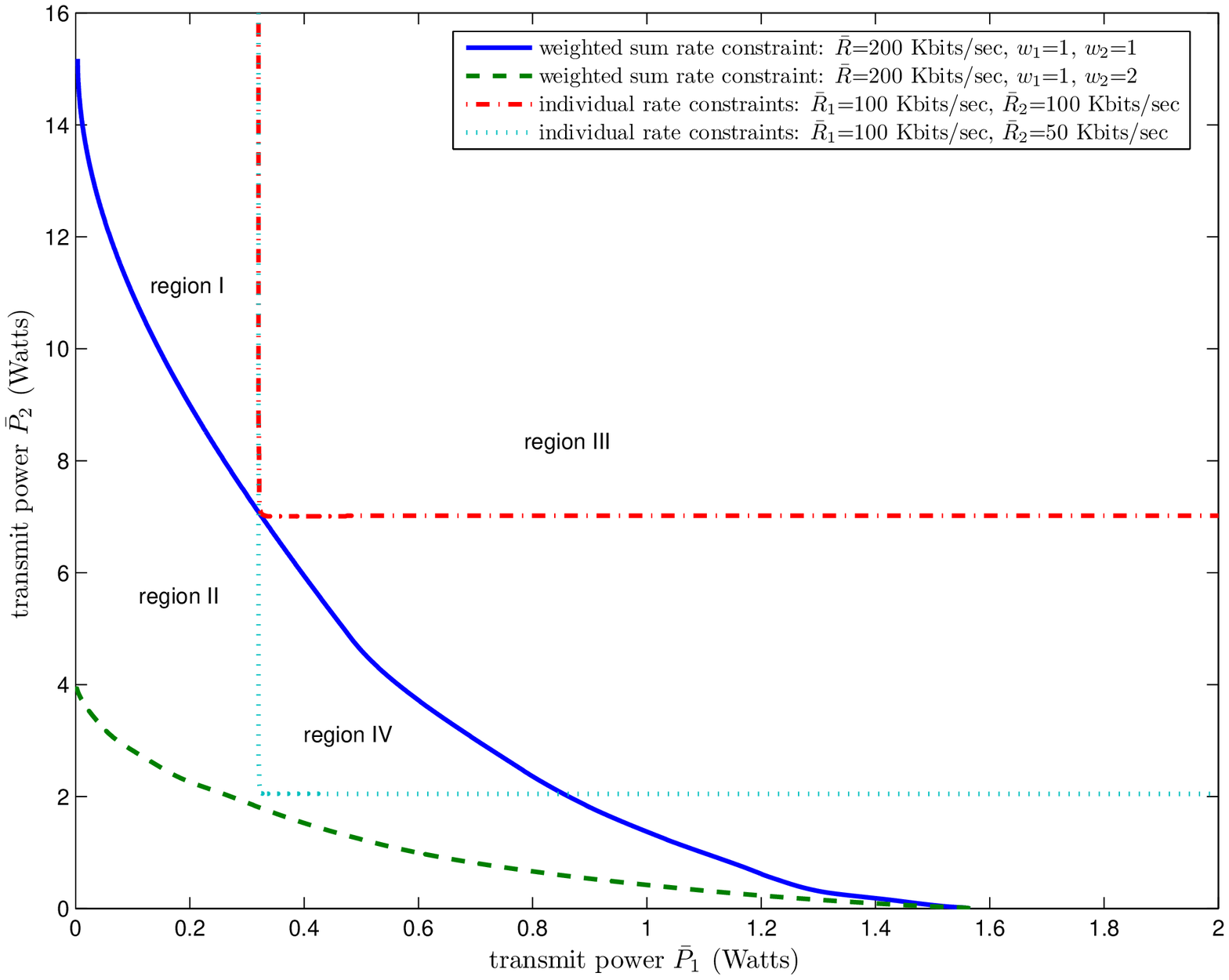, width=0.7\textwidth} \caption{
Power regions for the finite-AMC-mode case when two users have 10
dB difference in SNRs: $\bar{h}_1/(N_0B)=10$ dBW, and
$\bar{h}_2/(N_0B)=0$ dBW. } \label{F:pr4}
\end{figure}

Supposing $\bar{h}_1/(N_0B)=10$ dBW and $\bar{h}_1/(N_0B)=0$ dBW,
we also test our energy-efficient resource allocation under the
same four sets of rate constraints i)-iv). The power regions for
the infinite-codebook case and the finite-AMC-mode case are shown
in Fig.\ \ref{F:pr3} and Fig.\ \ref{F:pr4}, respectively. Since
the first user has a significantly better channel (i.e., higher
SNR) than user 2, the required transmit power of user 1 is much
lower than that of user 2 most of the time. Except this, the
results are similar to those in Figs.\ \ref{F:pr1} and
\ref{F:pr2}.

\begin{figure}
\centering \epsfig{file=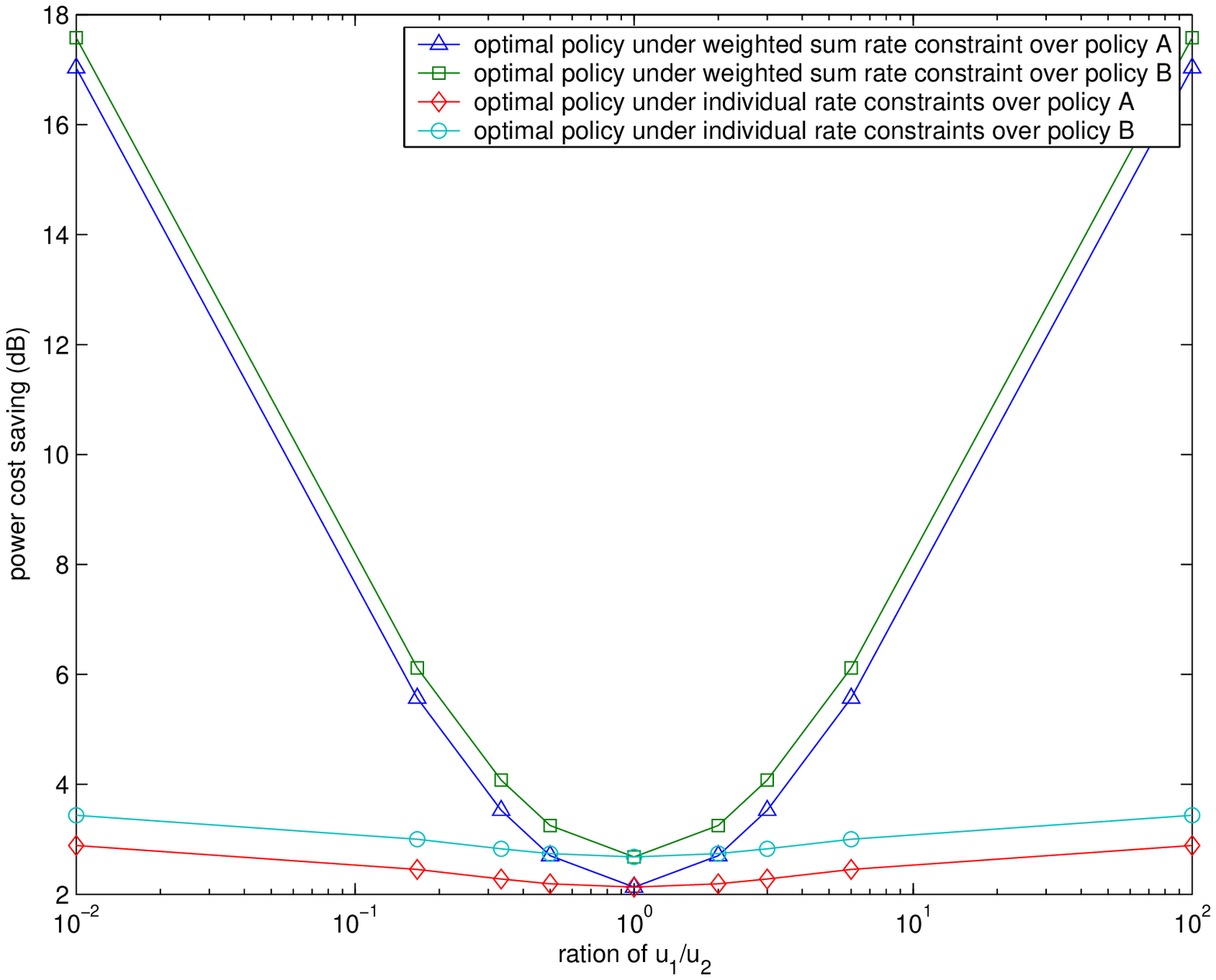, width=0.6\textwidth}
\caption{ Power savings for the infinite-codebook case when two
users have identical SNRs: $\bar{h}_1/(N_0B)=\bar{h}_2/(N_0B)=0$
dBW. (Policy A: equal time allocation and separate water-filling;
Policy B: equal time allocation among users and equal power per
fading state for each user.) } \label{F:comp1}
\end{figure}

\begin{figure}
\centering \epsfig{file=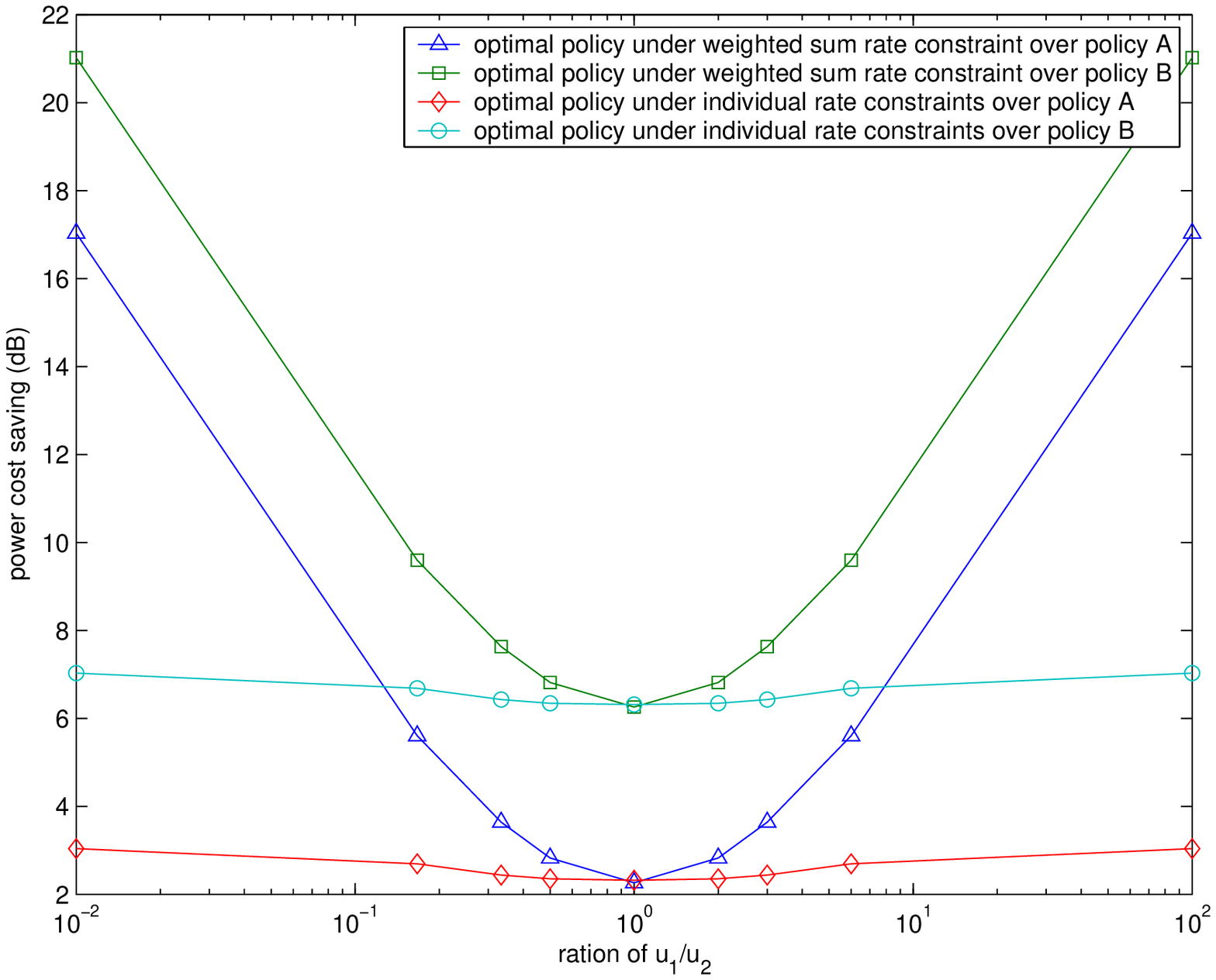, width=0.6\textwidth}
\caption{ Power savings for the finite-AMC-mode case when two
users have identical SNRs: $\bar{h}_1/(N_0B)=\bar{h}_2/(N_0B)=0$
dBW. (Policy A: equal time allocation and separate water-filling;
Policy B: equal time allocation among users and equal power per
fading state for each user.) } \label{F:comp2}
\end{figure}

\begin{figure}
\centering \epsfig{file=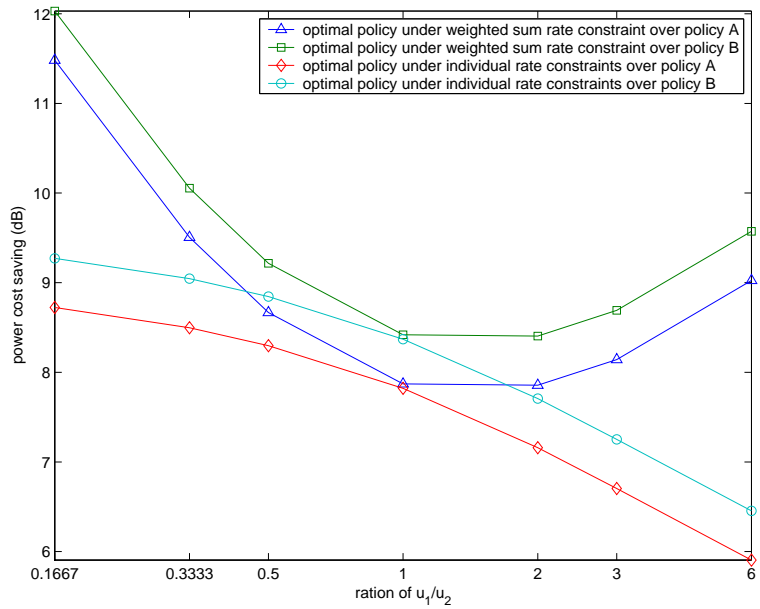, width=0.6\textwidth}
\caption{ Power savings for the infinite-codebook case when two
users have 10 dB difference in SNRs: $\bar{h}_1/(N_0B)=10$ dBW,
and $\bar{h}_2/(N_0B)=0$ dBW. (Policy A: equal time allocation and
separate water-filling; Policy B: equal time allocation among
users and equal power per fading state for each user.) }
\label{F:comp3}
\end{figure}

We next compare the derived energy-efficient resource allocation
with two alternative resource allocation policies. Policy A
assigns equal time fractions to the two users per block. Then each
user implements water-filling separately to adapt its transmission
rate at each assigned time fraction. In policy B, each user is
assigned equal time fraction and transmits with equal power per
block. Fig.\ \ref{F:comp1} depicts the power savings of our
optimal policies under two different sets of rate constraints i)
and iii), over the policies A and B for the infinite-codebook case
when two users have identical SNRs. It is seen that when the ratio
of two users' power cost weights is far away 1, our optimal
policies under a weighted sum-rate constraint can result in huge
power savings (near 20 dB) over the other two sub-optimal polices.
However, in this case the optimal policies under individual rate
constraints only have a small advantage (around 3 dB) in power
savings, over the sub-optimal policies. This is because with the
weighted sum average rate constraint, we can employ more flexible
policies in time and rate allocations. From Fig.\ \ref{F:comp1},
we also observe that the separate water-filling in policy A only
achieves small power savings (less than 1 dB) over the equal power
strategy in policy B. Fig.\ \ref{F:comp2} depicts the same
comparison for the finite-AMC-mode case. The same trends are
observed. However, in this case separate water-filling in policy A
achieves considerable power savings (4 dB) over the equal power
strategy in policy B. Fig.\ \ref{F:comp3} depicts similar power
savings for the infinite-codebook case under two different sets of
rate constraints ii) and iv), when two users have 10 dB in SNR
difference. Similar observations are obtained. But note that the
optimal policies under individual rate constraints can also
achieve large power savings (near 9 dB), over the sub-optimal
policies. In a nutsell, our energy-efficient resource allocation
policies may indeed result in large power savings.

\section{Concluding Remarks}\label{S:conc}

Given full or quantized CSI at the transmitters, we derived
energy-efficient resource allocation strategies for TDMA fading
channels. For energy minimization under a weighted average
sum-rate constraint, the optimal allocation policies are given by
water-filling over realizations of convex envelopes; whereas for
energy minimization under average individual rate constraints, the
optimal strategies perform greedy water-filling. Comparing these
two strategies for two different optimizations, we find that the
first approach requires one to characterize the convex envelope of
the minima of C$\thicksim$R functions per fading state, but the
associated scalar Lagrange multiplier $\lambda^*$ can be easily
obtained by one-dimensional search. Greedy water-filling simply
computes and compares the channel quality indicator functions of
individual users and then picks the user(s) with best channel(s)
to transmit per block; however, we need to iteratively compute the
associated vector Lagrange multiplier $\boldsymbol{\lambda}^*$ (by
Algorithm 4).

An interesting feature of our energy-efficient resource allocation
strategies should be stressed. According to our policies, we can
let the access point (which naturally has full CSI) decide the
time allocation and feed it back to users via uplink map messages,
as in e.g., IEEE 802.16 systems~\cite{80216}. Then given the
Lagrange constant $\lambda^*$ (with the weighted sum-rate
constraint) or vector $\boldsymbol{\lambda}^*$ (with the
individual rate constraints), the user only needs its own CSI to
determine the transmission rate at the assigned time fraction. If
uplink and downlink transmissions are operated in a time-division
duplexing (TDD) mode, the users can even obtain their own CSI
without feedback from the access point. Together with the fact
that the access point needs only a few bits to indicate the time
allocation (since most of the time we should allow only one user
to transmit), this feature is attractive from a practical
implementation viewpoint.

As far as future work, it is interesting to study energy
minimization over fading channels with delay constraint and/or
using quantized CSI throughout. Our energy minimization for
finite-AMC modes only requires quantized CSI feedback from the
access point. And delay-constrained energy minimization may be
seen as the dual problem of the delay-limited capacity
maximization in~\cite{Hanly&Tse1998}. Extensions to these two
directions are currently under investigation.\footnote{The views
and conclusions contained in this document are those of the
authors and should not be interpreted as representing the official
policies, either expressed or implied, of the Army Research
Laboratory or the U. S. Government.}

\section*{Appendices}

\subsection{Proof of Lemmas 1 and 2}

\emph{A1. Proof of Lemma 1:} The $k$th derivatives of $f_2(x)$ and
$f_1(x)$ are given by
\begin{equation}
    f_2^{(k)}(x)=\frac{(\ln 2)^k \mu_2}{w_2^k h_2} 2^{\frac{x}{w_2}}, \qquad
    f_1^{(k)}(x)=\frac{(\ln 2)^k \mu_1}{w_1^k h_1} 2^{\frac{x}{w_1}}.
\end{equation}

\begin{enumerate}

\item If $\frac{\mu_1}{w_1h_1} \geq \frac{\mu_2}{w_2h_2}$, since
$2^{\frac{x}{w_2}} < 2^{\frac{x}{w_1}}$ for $w_1 < w_2$, we have
$f_2^{(1)}(x)<f_1^{(1)}(x)$, $\forall x > 0$. Since
$f_2(0)=f_1(0)$, we readily infer that $f_2(x)<f_1(x)$, $\forall x
> 0$.

\item If $\frac{\mu_1}{w_1h_1} < \frac{\mu_2}{w_2h_2}$, since $w_1
< w_2$, there exists $k>1$ such that $\frac{(\ln 2)^k \mu_2}{w_2^k
h_2} \leq \frac{(\ln 2)^k \mu_1}{w_1^k h_1}$. Let $k_0 = \min_{k}
\; \arg \left\{\frac{(\ln 2)^k \mu_2}{w_2^k h_2} \leq \frac{(\ln
2)^k \mu_1}{w_1^k h_1}\right\}$. Together with $2^{\frac{x}{w_2}}
< 2^{\frac{x}{w_1}}$, we have $f_2^{(k_0)}(x)<f_1^{(k_0)}(x)$,
$\forall x \geq 0$. Since $f_2^{(k_0-1)}(0)>f_1^{(k_0-1)}(0)$ and
$f_2^{(k_0)}(x)<f_1^{(k_0)}(x)$, we infer that
$f_2^{(k_0-1)}(x)-f_1^{(k_0-1)}(x)$ starts positive and with a
negative slope it crosses the $x$-axis at some point $v_{k_0-1}$;
hence,
\begin{equation}\label{E:f0k}
    \begin{cases}
        f_2^{(k_0-1)}(x) > f_1^{(k_0-1)}(x) & \mbox{when } 0 \leq x < v_{k_0-1}, \\
        f_2^{(k_0-1)}(x) < f_1^{(k_0-1)}(x) & \mbox{when } x > v_{k_0-1};
    \end{cases}
\end{equation}
Using (\ref{E:f0k}) and $f_2^{(k_0-2)}(0)>f_1^{(k_0-2)}(0)$, we
obtain similar results for $f_2^{(k_0-2)}(x)$ and
$f_1^{(k_0-2)}(x)$ with $v_{k_0-2} > v_{k_0-1}$. By induction, we
therefore deduce that
\begin{equation}
    \begin{cases}
        f_2(x) > f_1(x) & \mbox{when } 0 \leq x < v_0, \\
        f_2(x) < f_1(x) & \mbox{when } x > v_0;
    \end{cases}
\end{equation}
where $v_0$ is the unique solution of the equation
$f_2(x)=f_1(x)$.
\end{enumerate}

\emph{A2. Proof of Lemma 2:} For notational brevity, we drop the
dependence of $R$, $R_k$, $\tau_k$ ($k=1,2$), $R_a$, $R_b$, and
$s_0$ on $\mathbf{h}$. And we let $\tau_1=\tau$ and
$\tau_2=1-\tau$.
\begin{enumerate}

\item When $w_1<w_2$ and $\frac{\mu_1}{w_1h_1} \geq
\frac{\mu_2}{w_2h_2}$, we have from Lemma 1 $f_2(x)<f_1(x)$,
$\forall x > 0$. We wish to solve (\ref{E:JR}) under the
constraint $\tau R_1+(1-\tau) R_2 =R$ with $R_1 \geq 0$ and $R_2
\geq 0$. Then the cost function satisfies
\begin{equation}\label{E:eq69}
    \tau f_1(R_1)+(1-\tau) f_2(R_2) \geq \tau f_2(R_1)+(1-\tau)
    f_2(R_2) \geq f_2(R),
\end{equation}
where the last inequality is due to the convexity of $f_2(R)$, and
the equalities are achieved when $\tau=0$, $R_1=0$ and $R_2=R$.
Inequality (\ref{E:eq69}) clearly shows that the minimum in
(\ref{E:JR}) is achieved; i.e., $J(R)=f_2(R)$.

\item When $w_1 < w_2$ and $\frac{\mu_1}{w_1h_1} <
\frac{\mu_2}{w_2h_2}$, $f_1(R)$ and $f_2(R)$ intersect as depicted
in Fig.\ \ref{F:crf}. Similar to \cite[Proof of Lemma
1]{Li&Goldsmith2001a}, we can specify two points $R_a$ and $R_b$
such that $f_1^{(1)}(R_a)=f_2^{(1)}(R_b)=s_0$; i.e., at these two
points first-order derivatives of $f_1(R)$ and $f_2(R)$ are equal:
\begin{equation}\label{E:Rab}
    f_1^{(1)}(R_a)=\frac{\ln 2 \mu_1}{w_1 h_1} 2^{\frac{R_a}{w_1}} =
    s_0; \quad \quad
    f_2^{(1)}(R_b)=\frac{\ln 2 \mu_2}{w_2 h_2} 2^{\frac{R_b}{w_2}} =
    s_0.
\end{equation}
The $R_a$, $R_b$ expressions are obtained after solving
(\ref{E:Rab}) for $R_a$ and $R_b$. Since $s_0$ is the slope of the
common tangent line of the curves $f_1(R)$ and $f_2(R)$, we also
have
\begin{equation}\label{E:gamma0}
    s_0 = \frac{f_2(R_b)-f_1(R_a)}{R_b-R_a}.
\end{equation}
Solving (\ref{E:Rab}) and (\ref{E:gamma0}), we obtain $s_0$ as the
solution to $g(x,\mathbf{h})=0$, where $g(x,\mathbf{h})$ is given
by (\ref{E:gx}), and $R_a$ and $R_b$ are as in (\ref{E:Rabh}). If
$0 < R < R_a$ or $R > R_b$, then $J(R)$ simply equals to $f_1(R)$
or $f_2(R)$. If $R_a < R < R_b$, then $J(R)$ should take the
values between $f_1(R_a)$ and $f_2(R_b)$ on the straight line, and
$J(R)$ can be achieved by time-sharing; i.e., $R=\tau R_a +
(1-\tau) R_b$, and $J(R)=\tau f_1(R_a) + (1-\tau) f_2(R_b)$. The
optimal resource allocation per fading state is thus given by
(\ref{E:rt1})-(\ref{E:rt3}).
\end{enumerate}

\subsection{Proof of Theorem 1}

Let $R^*(\mathbf{h})$ denote the optimal total rate reward
assigned to fading state $\mathbf{h}$, and $r_k^*(\mathbf{h})$ and
$\tau_k^*(\mathbf{h})$ the corresponding optimal rate and time
fraction allocated to user $k=1,2$. To solve (\ref{E:wf}), we rely
on the Karush-Kuhn-Tucker (KKT) conditions~\cite[Chapter
5]{convex}. Let $J^{(1)}(R(\mathbf{h}))$ denote first derivative
of $J(R(\mathbf{h}))$ with respect to $R(\mathbf{h})$. Taking the
partial derivative of $E_{\mathbf{h}} [J(R(\mathbf{h}))] - \lambda
E_{\mathbf{h}}[R(\mathbf{h})]$ in (\ref{E:wf}) with respect to
$R(\mathbf{h})$ for a fixed $\mathbf{h}$, only
$J^{(1)}(R(\mathbf{h})) - \lambda$ survives since all
$R(\mathbf{h'})$ for realizations $\mathbf{h'} \neq \mathbf{h}$
are regarded as constants. We thus have at the optimum that if
$R^*(\mathbf{h})>0$,
\begin{equation}\label{E:JpR}
    J^{(1)}(R^*(\mathbf{h})) = \lambda^*.
\end{equation}

\begin{enumerate}
\item If $\frac{\mu_1}{w_1h_1} \geq \frac{\mu_2}{w_2h_2}$, then
$J(R(\mathbf{h})) =f_2(R(\mathbf{h}))$ from Lemma 2, and thus
$J^{(1)}(R(\mathbf{h}))=\frac{\ln 2 \mu_2}{w_2 h_2}
2^{\frac{R(\mathbf{h})}{w_2}}$. Substituting the latter into
(\ref{E:JpR}) and recalling that $R(\mathbf{h}) \geq 0$, we find
\begin{equation}
    R^*(\mathbf{h}) = w_2 \left[\log \frac{\lambda^* w_2
    h_2}{\ln 2 \mu_2}\right]_+ = w_2 \left[\log \lambda^* -
    \log \frac{\ln 2 \mu_2}{w_2h_2}\right]_+,
\end{equation}
which yields the optimal rate and time allocation in
(\ref{E:2U1}).

\item If $\frac{\mu_1}{w_1h_1} < \frac{\mu_2}{w_2h_2}$ and we
recall that $s_0$ is the slope of the straight line segment in
Fig.\ \ref{F:crf}, it follows from Lemma 2 that
\begin{equation}\label{E:eq73}
    \begin{cases}
        \mbox{if} \;\; 0 < R^*(\mathbf{h}) < R_a, &
        \lambda^*=J^{(1)}(R^*(\mathbf{h})) < s_0; \\
        \mbox{if} \;\; R_a \leq R^*(\mathbf{h}) \leq R_b, &
        \lambda^*=J^{(1)}(R^*(\mathbf{h})) = s_0; \\
        \mbox{if} \;\; R^*(\mathbf{h}) > R_b, &
        \lambda^*=J^{(1)}(R^*(\mathbf{h})) > s_0;
    \end{cases}
\end{equation}
where $R_a$, $R_b$ and $s_0$ are functions of $\mathbf{h}$, given
in Lemma 2. Now using (\ref{E:eq73}), we arrive at:

\begin{enumerate}
\item If $\lambda^* < s_0$, then $J(R(\mathbf{h}))
=f_1(R(\mathbf{h}))$, and thus $R^*(\mathbf{h}) = w_1 \left[\log
\lambda^* - \log \frac{\ln 2 \mu_1}{w_1h_1}\right]_+$; which in
turn yields the optimal rate and time allocation in (\ref{E:2U2}).

\item If $\lambda^* > s_0$, then $J(R(\mathbf{h}))
=f_2(R(\mathbf{h}))$, and thus $R^*(\mathbf{h}) = w_2 \left[\log
\lambda^* - \log \frac{\ln 2 \mu_2}{w_2h_2}\right]_+$, which in
turn yields the optimal rate and time allocation in (\ref{E:2U3}).

\item If $\lambda^* = s_0$, then $J(R(\mathbf{h})) =f_1(R_a)+s_0
(R(\mathbf{h})-R_a)$, and $R^*(\mathbf{h})$ can be any value
between $R_a$ and $R_b$. In this case, we obtain the optimal rate
and time allocation (\ref{E:2U4}).
\end{enumerate}

Since $R_a < R_b$, we have from (\ref{E:Rabh})
\begin{equation}\label{E:eq75}
    w_1 \log \frac{s_0w_1h_1}{\ln 2\mu_1} < w_2 \log
    \frac{s_0w_2h_2}{\ln 2\mu_2} \quad \Rightarrow \quad s_0
    > \left(\frac{((w_1h_1)/(\ln 2\mu_1))^{w_1}}{((w_2h_2)/(\ln 2\mu_2))^{w_2}}
    \right)^{\frac{1}{w_2-w_1}}:=\xi.
\end{equation}
Moreover, since $g(s_0,\mathbf{h})=0$, and
$g^{(1)}(x,\mathbf{h})=w_2\log \frac{xw_2h_2}{\ln 2\mu_2}-w_1\log
\frac{xw_1h_1}{\ln 2\mu_1}$, it follows from (\ref{E:eq75}) that
\begin{equation}
    \begin{cases}
        g^{(1)}(x,\mathbf{h}) < 0, & \mbox{for} \;\; 0 < x < \xi;
        \\
        g^{(1)}(x,\mathbf{h}) > 0, & \mbox{for} \;\; x > \xi.
    \end{cases}
\end{equation}
Therefore, we have:
\begin{enumerate}
\item[a')] If $\lambda^* < \xi$ or $\lambda^* > \xi$ and
$g(\lambda^*,\mathbf{h})<0$, then $\lambda^* < s_0$.

\item[b')] If $\lambda^* > \xi$ and $g(\lambda^*,\mathbf{h})>0$,
then $\lambda^* > s_0$.

\item[c')] If $\lambda^* > \xi$ and $g(\lambda^*,\mathbf{h})=0$,
then $\lambda^* = s_0$.
\end{enumerate}
Combining a'), b'), c') with a), b), c) proves the second part of
Theorem 1.
\end{enumerate}

\subsection{Proof of Theorem 2}

Since $\bar{f}(x):=\min_{1 \leq k \leq K} f_k(x)$, it is easy to
show that
\begin{equation}\label{E:ce1}
    J(R(\mathbf{h})):= \min_{\mathbf{r}(\mathbf{h}),
        \boldsymbol{\tau}(\mathbf{h})} \sum_{k=1}^K \tau_k(\mathbf{h})
        f_k (R_k(\mathbf{h})) \geq \min_{\mathbf{r}(\mathbf{h}),
        \boldsymbol{\tau}(\mathbf{h})} \sum_{k=1}^K \tau_k(\mathbf{h})
        \bar{f} (R_k(\mathbf{h})).
\end{equation}
By the definition of the convex envelope $\bar{f}^c(x)$, we have
\begin{equation}\label{E:ce2}
    \sum_{k=1}^K \tau_k(\mathbf{h}) \bar{f}
    (R_k(\mathbf{h})) \geq \sum_{k=1}^K \tau_k(\mathbf{h})
    \bar{f}^c (R_k(\mathbf{h})) \geq \bar{f}^c
    \left(\sum_{k=1}^K
    \tau_k(\mathbf{h})R_k(\mathbf{h})\right) :=
    \bar{f}^c (R(\mathbf{h}));
\end{equation}
where the last inequality is due to the convexity of
$\bar{f}^c(x)$. From (\ref{E:ce1}) and (\ref{E:ce2}), it is clear
that $J(R(\mathbf{h})) \geq \bar{f}^c (R(\mathbf{h}))$.

On the other hand, according to the definition of $\bar{f}^c(x)$,
for any given $R(\mathbf{h})$ there exists $R_a(\mathbf{h})$ and
$R_b(\mathbf{h})$ (possibly $R_a(\mathbf{h})=R_b(\mathbf{h})$) and
$\tau^*$ such that
\begin{equation}\label{E:eq80}
    \tau^* R_a(\mathbf{h}) + (1-\tau^*) R_b(\mathbf{h}) =
    R(\mathbf{h}); \quad \quad \tau^* \bar{f}(R_a(\mathbf{h})) +
    (1-\tau^*) \bar{f}(R_b(\mathbf{h})) =
    \bar{f}^c(R(\mathbf{h})).
\end{equation}
Eq. (\ref{E:eq80}) shows that we can achieve the equality
$J(R(\mathbf{h})) = \bar{f}^c (R(\mathbf{h}))$ by assigning
\begin{equation}
    \begin{cases}
        \tau_i^* (\mathbf{h}) = \tau^*, & r_i^* (\mathbf{h}) =
        R_a(\mathbf{h})/w_i \\
        \tau_j^* (\mathbf{h}) = 1-\tau^*, & r_j^* (\mathbf{h}) =
        R_b(\mathbf{h})/w_j
    \end{cases}
\end{equation}
to users $i$ and $j$ which satisfy
$f_i(R_a(\mathbf{h}))=\bar{f}(R_a(\mathbf{h}))$ and
$f_j(R_b(\mathbf{h}))=\bar{f}(R_b(\mathbf{h}))$. This completes
the proof.

\subsection{Proof of Theorem 3}

Letting $M:=\sum_{k=1}^{K} M_k$, the value of $K_0$ per fading
state may take any integer in [1, $M$]. We define $\mathbf{{\cal
H}}_{K_0}$ as the set of all fading states $\mathbf{h}$ for which
Algorithm 3 yields $K_0 \in [1, M]$. We further set $s_{K_0+1}
(\mathbf{h}) = \infty$, $\forall \lambda
> 0$, and define for $m=0,1,\ldots, K_0$, the sets
\begin{eqnarray}
    \mathbf{{\cal H}}_m (K_0, \lambda) & :=  & \left\{\mathbf{h}:\; \mathbf{h} \in
    \mathbf{{\cal H}}_{K_0}, \; s_m (\mathbf{h}) < \lambda < s_{m+1}
    (\mathbf{h})\right\}, \\
    \tilde{\mathbf{{\cal H}}}_m (K_0, \lambda) & :=  & \left\{\mathbf{h}:\; \mathbf{h} \in
    \mathbf{{\cal H}}_{K_0}, \; \lambda = s_m (\mathbf{h}) \right\}.
\end{eqnarray}
Then $\forall \lambda > 0$, since $\tilde{\mathbf{{\cal H}}}_0
(K_0, \lambda) = \phi$, we can express the set $\mathbf{{\cal H}}$
of all possible fading states as
\begin{equation}
    \mathbf{{\cal H}} = \bigcup_{1 \leq K_0 \leq M} \; \mathbf{{\cal H}}_{K_0} = \bigcup_{1 \leq K_0 \leq
    M} \; \left\{\mathbf{{\cal H}}_0 (K_0, \lambda) \cup \bigcup_{1 \leq m \leq K_0} \; \left(\mathbf{{\cal H}}_m (K_0, \lambda) \cup \tilde{\mathbf{{\cal H}}}_m (K_0,
    \lambda)\right)\right\}.
\end{equation}
Using the defined partitions $\mathbf{{\cal H}}_{K_0}$, we can
rewrite (\ref{E:optimal3}) as
\begin{equation}
    \begin{cases}
        \min_{\boldsymbol{\tilde{\tau}}(\mathbf{h})} \; \sum_{K_0=1}^{M} \; E_{\mathbf{h} \in \mathbf{{\cal H}}_{K_0}}
        \left[\sum_{m=1}^{K_0} \tilde{\tau}_m(\mathbf{h})
        C_m(\mathbf{h})\right] \\
        \mbox{subject to} \quad \quad \sum_{K_0=1}^{M} \; E_{\mathbf{h} \in \mathbf{{\cal H}}_{K_0}}
        \left[\sum_{m=1}^{K_0} \tilde{\tau}_m(\mathbf{h})
        R_m(\mathbf{h})\right] \geq \bar{R}, \quad
        \sum_{m=1}^{K_0} \tilde{\tau}_m(\mathbf{h}) = 1.
    \end{cases}
\end{equation}

If the optimization is feasible, we have
\begin{equation}
    \sum_{K_0=1}^M  \; E_{\mathbf{h} \in \mathbf{{\cal H}}_{K_0}}
    \left[ R_{K_0}(\mathbf{h}) \right] \; \geq \bar{R}.
\end{equation}
It is easy to show that we can always achieve (\ref{E:achieve})
although the solution for $\tau_0^*$ may not be unique.

In the following, $\forall K_0 \in [1, M]$ and $\forall m \in [1,
K_0]$, we drop the dependence of $\tilde{\tau}_m$,
$\tilde{\tau}_m^*$, $R_m$, $C_m$ and $s_m$ on $\mathbf{h}$ for
notational brevity. For $K_0 \in [1, M]$ and $\forall \mathbf{h}
\in \mathbf{{\cal H}}_{K_0}$, we let
\begin{equation}
    \boldsymbol{\tilde{\tau}}^*:=[\tilde{\tau}_1^*, \ldots, \tilde{\tau}_{K_0}^*]^T, \quad \quad
    \boldsymbol{\tilde{\tau}}:=[\tilde{\tau}_1, \ldots, \tilde{\tau}_{K_0}]^T.
\end{equation}
The $\boldsymbol{\tilde{\tau}}^*$ given by the theorem satisfies
\begin{equation}
    \sum_{K_0=1}^M \; E_{\mathbf{h} \in \mathbf{{\cal H}}_{K_0}} \left[
    \sum_{m=1}^{K_0} \tilde{\tau}_m^* R_m \right] =  \bar{R}.
\end{equation}
From the definition of $\{\mathbf{{\cal H}}_m(K_0,
\lambda^*)\}_{m=0}^{K_0}$ and $\{\tilde{\mathbf{{\cal H}}}_m(K_0,
\lambda^*)\}_{m=1}^{K_0}$, we have $\forall
\boldsymbol{\tilde{\tau}} \neq \boldsymbol{\tilde{\tau}}^*$ that
\begin{eqnarray}
    \nonumber
    \lefteqn{\sum_{K_0=1}^M \; E_{\mathbf{h} \in \mathbf{{\cal H}}_{K_0}} \left[
    \sum_{m=1}^{K_0} \tilde{\tau}_m R_m \right] - \bar{R}} \\
    \nonumber
    & = &
    \sum_{K_0=1}^M \; E_{\mathbf{h} \in \mathbf{{\cal H}}_{K_0}} \left[
    \sum_{m=1}^{K_0} \tilde{\tau}_m R_m \right] - \sum_{K_0=1}^M \; E_{\mathbf{h} \in \mathbf{{\cal H}}_{K_0}} \left[
    \sum_{m=1}^{K_0} \tilde{\tau}_m^* R_m \right] \\
    \nonumber
    & = & \sum_{K_0=1}^M \; \int_{\mathbf{h} \in \mathbf{{\cal H}}_0(K_0, \lambda^*)}
    \sum_{i=1}^{K_0} \tilde{\tau}_i R_i \; dF(\mathbf{h}) \\
    \nonumber
    & & + \sum_{K_0=1}^M \; \sum_{m=1}^{K_0} \int_{\mathbf{h} \in \mathbf{{\cal H}}_m(K_0,\lambda^*)}
    \left(\sum_{i=1}^{K_0} \tilde{\tau}_i R_i -R_m\right) \; dF(\mathbf{h}) \\
    & & + \sum_{K_0=1}^M \; \sum_{m=1}^{K_0} \int_{\mathbf{h} \in \tilde{\mathbf{{\cal H}}}_m(K_0,\lambda^*)}
    \left(\sum_{i=1}^{K_0} \tilde{\tau}_i R_i -\tau_0^*R_m - (1-\tau_0^*)R_{m-1}\right) \;
    dF(\mathbf{h}).
    \label{E:difference}
\end{eqnarray}
By the convexity of $J(R(\mathbf{h}))$, we can easily show that
\begin{equation}\label{E:slope}
    \frac{C_{m-1}}{R_{m-1}} < \frac{C_m}{R_m},
    \quad \quad m=2,\ldots, K_0.
\end{equation}

\begin{enumerate}
\item $\forall \mathbf{h} \in \mathbf{{\cal H}}_0(K_0,
\lambda^*)$, since $\lambda^* < s_1 = C_1/R_1$, we have from
(\ref{E:slope}) that $\lambda^* < C_i/R_i$, $\forall i \in [1,
K_0]$, i.e., $R_i < C_i/\lambda^*$. Then the first summand of the
last equality in (\ref{E:difference}) will be
\begin{equation}\label{E:A1}
    \sum_{K_0=1}^M \; \int_{\mathbf{h} \in \mathbf{{\cal H}}_0(K_0, \lambda^*)}
    \sum_{i=1}^{K_0} \tilde{\tau}_i R_i \; dF(\mathbf{h}) \leq \frac{1}{\lambda^*}
    \sum_{K_0=1}^M
    \; \int_{\mathbf{h} \in \mathbf{{\cal H}}_0(K_0, \lambda^*)} \sum_{i=1}^{K_0} \tilde{\tau}_i C_i \;
    dF(\mathbf{h}).
\end{equation}

\item $\forall \mathbf{h} \in \mathbf{{\cal H}}_m(K_0,\lambda^*)$,
$m=1,\ldots, K_0$, since $s_m < \lambda^* < s_{m+1}$, we have
\begin{equation}
    R_{m+1} < R_m + (C_{m+1}-C_m)/\lambda^*, \quad \quad
    R_{m-1} < R_m + (C_{m-1}-C_m)/\lambda^*.
\end{equation}
Using convexity, we have $s_i < \lambda^* < s_k$, $\forall i \in
[1, m-1]$, $\forall k \in [m+1, K_0]$. Hence $\forall i \neq m$,
$i \in [1, K_0]$, we have $R_i < R_m + (C_i-C_m)/\lambda^*$. Then
the second summand in (\ref{E:difference}) will be
\begin{eqnarray}
    \nonumber
    \lefteqn{\sum_{K_0=1}^M \; \sum_{m=1}^{K_0} \int_{\mathbf{h} \in \mathbf{{\cal H}}_m(K_0,\lambda^*)}
    \left(\sum_{i=1}^{K_0} \tilde{\tau}_i R_i -R_m\right) \;
    dF(\mathbf{h})} \\
    \nonumber
    & \leq & \sum_{K_0=1}^M \; \sum_{m=1}^{K_0} \int_{\mathbf{h} \in
    \mathbf{{\cal H}}_m(K_0,\lambda^*)} \left(\sum_{i\neq m, \; i=1}^{K_0} \tilde{\tau}_i [R_m +
    (C_i-C_m)/\lambda^*] \; +\tilde{\tau}_m
    R_m-R_m\right)\;dF(\mathbf{h})\\
    \nonumber
    & = & \sum_{K_0=1}^M \; \sum_{m=1}^{K_0} \int_{\mathbf{h} \in
    \mathbf{{\cal H}}_m(K_0,\lambda^*)} \left(\sum_{i\neq m, \; i=1}^{K_0} \tilde{\tau}_i
    (C_i-C_m)/\lambda^*\right)\;dF(\mathbf{h}) \\
    & = & \frac{1}{\lambda^*} \sum_{K_0=1}^M \; \sum_{m=1}^{K_0} \int_{\mathbf{h} \in
    \mathbf{{\cal H}}_m(K_0,\lambda^*)} \left(\sum_{i=1}^{K_0}
    \tilde{\tau}_i C_i-C_m\right)\;dF(\mathbf{h}).
    \label{E:A2}
\end{eqnarray}

\item $\forall \mathbf{h} \in \tilde{\mathbf{{\cal
H}}}_m(K_0,\lambda^*)$, $m \in [1, K_0]$ and since $\lambda^* =
s_m$, we have $R_{m-1} = R_m + (C_{m-1}-C_m)/\lambda^*$.
Since $\forall i \in [1, m-1]$, $\forall k \in [m+1,K_0]$, we have
$s_i < \lambda^* < s_k$, from which it follows that $R_i < R_m +
(C_i-C_m)/\lambda^*$, $\forall i \neq m, m-1$ and $i \in [1,
K_0]$. Then the third summand in (\ref{E:difference}) will be
\begin{eqnarray}
    \nonumber
    \lefteqn{\sum_{K_0=1}^M \; \sum_{m=1}^{K_0} \int_{\mathbf{h} \in \tilde{\mathbf{{\cal H}}}_m(K_0,\lambda^*)}
    \left(\sum_{i=1}^{K_0} \tilde{\tau}_i R_i -\tau_0^*R_m - (1-\tau_0^*)R_{m-1}\right) \;
    dF(\mathbf{h})} \\
    \nonumber
    & \leq & \sum_{K_0=1}^M \; \sum_{m=1}^{K_0} \int_{\mathbf{h} \in
    \tilde{\mathbf{{\cal H}}}_m(K_0,\lambda^*)} \left(\sum_{i\neq m, \; i=1}^{K_0} \tilde{\tau}_i [R_m +
    (C_i-C_m)/\lambda^*] \right. \\
    \nonumber
    & & \hspace{4.5cm} \Biggl. +\tilde{\tau}_m
    R_m-\tau_0^*R_m - (1-\tau_0^*)R_{m-1}\Biggr)\;dF(\mathbf{h}) \\ 
    \nonumber
    & = & \sum_{K_0=1}^M \; \sum_{m=1}^{K_0} \int_{\mathbf{h} \in
    \tilde{\mathbf{{\cal H}}}_m(K_0,\lambda^*)} \left(\sum_{i\neq m, \; i=1}^{K_0} \tilde{\tau}_i
    (C_i-C_m)/\lambda^* \; + (1-\tau_0^*) (R_m-R_{m-1})\right)\;dF(\mathbf{h}) \\
    & = & \frac{1}{\lambda^*} \sum_{K_0=1}^M \; \sum_{m=1}^{K_0} \int_{\mathbf{h} \in
    \tilde{\mathbf{{\cal H}}}_m(K_0,\lambda^*)} \left(\sum_{i=1}^{K_0}
    \tilde{\tau}_iC_i- \tau_0^*C_m - (1-\tau_0^*)C_{m-1}\right)\;dF(\mathbf{h}).
    \label{E:A3}
\end{eqnarray}

\end{enumerate}

Substituting (\ref{E:A1}), (\ref{E:A2}) and (\ref{E:A3}) into
(\ref{E:difference}), we find
\begin{eqnarray}
    \nonumber
    \lefteqn{\sum_{K_0=1}^M \; E_{\mathbf{h} \in \mathbf{{\cal H}}_{K_0}} \left[
    \sum_{m=1}^{K_0} \tilde{\tau}_m R_m \right] - \bar{R} }\\
    \nonumber
    & \leq &
    \frac{1}{\lambda^*} \sum_{K_0=1}^M \left\{ \int_{\mathbf{h} \in \mathbf{{\cal H}}_0(K_0,\lambda^*)}
    \sum_{i=1}^{K_0} \tilde{\tau}_i C_i \;
    dF(\mathbf{h})  \right. \\
    \nonumber
    & &  \hspace{2cm} +\sum_{m=1}^{K_0} \int_{\mathbf{h} \in
    \mathbf{{\cal H}}_m(K_0,\lambda^*)} \left(\sum_{i=1}^{K_0}
    \tilde{\tau}_iC_i-C_m\right)\;dF(\mathbf{h}) \\
    \nonumber
    & &  \hspace{2cm} \left. + \sum_{m=1}^{K_0} \int_{\mathbf{h} \in
    \tilde{\mathbf{{\cal H}}}_m(K_0,\lambda^*)} \left(\sum_{i=1}^{K_0}
    \tilde{\tau}_iC_i- \tau_0^*C_m -
    (1-\tau_0^*)C_{m-1}\right)\;dF(\mathbf{h})\right\}\\
    & = & \frac{1}{\lambda^*} \left\{ \sum_{K_0=1}^M \; E_{\mathbf{h} \in \mathbf{{\cal H}}_{K_0}} \left[
    \sum_{m=1}^{K_0} \tilde{\tau}_m C_m \right] - \sum_{K_0=1}^M \; E_{\mathbf{h} \in \mathbf{{\cal H}}_{K_0}} \left[
    \sum_{m=1}^{K_0} \tilde{\tau}_m^* C_m \right] \right\}.
\end{eqnarray}

Therefore, $\forall \boldsymbol{\tilde{\tau}} \neq
\boldsymbol{\tilde{\tau}}^*$, if $\boldsymbol{\tilde{\tau}}$
satisfies the total average-rate constraint $\sum_{K_0=1}^M \;
E_{\mathbf{h} \in \mathbf{{\cal H}}_{K_0}} \left[\sum_{m=1}^{K_0}
\tilde{\tau}_m R_m \right] \geq \bar{R}$, we have
\begin{equation}
    \sum_{K_0=1}^M \; E_{\mathbf{h} \in \mathbf{{\cal H}}_{K_0}} \left[
    \sum_{m=1}^{K_0} \tilde{\tau}_m C_m \right] \geq \sum_{K_0=1}^M \; E_{\mathbf{h} \in \mathbf{{\cal H}}_{K_0}} \left[
    \sum_{m=1}^{K_0} \tilde{\tau}_m^* C_m \right].
\end{equation}
Hence, $\boldsymbol{\tilde{\tau}}^*$ is the optimal solution to
(\ref{E:optimal3}) and consequently the corresponding
$\mathbf{r}^*$ and $\boldsymbol{\tau}^*$ are the optimal solutions
to (\ref{E:optimal2}).

\subsection{Proof of Theorem 4}

Following the Lagrange multiplier method, (\ref{E:ind2}) is
equivalent to
\begin{equation}\label{E:ind4}
    \begin{cases}
        \min_{\mathbf{r}(\cdot), \boldsymbol{\tau}(\cdot)} E_{\mathbf{h}} \left[ \sum_{k=1}^K \mu_k
        \frac{\tau_k(\mathbf{h})}{h_k} (2^{r_k(\mathbf{h})}-1)\right] -
        \sum_{k=1}^K \lambda_k E_{\mathbf{h}} [\tau_k(\mathbf{h})
        r_k(\mathbf{h})]
        \\
        \mbox{subject to} \quad E_{\mathbf{h}} \left[ \tau_k(\mathbf{h})
        r_k(\mathbf{h})\right] = \bar{R}_k, \;\; k=1,\ldots, K, \\
        \hspace{2cm}  \sum_{k=1}^K \tau_k(\mathbf{h}) =1, \;\; \forall
        \mathbf{h}.
    \end{cases}
\end{equation}
Using the definition of $f_k(r_k(\mathbf{h}))$, we can rewrite
(\ref{E:ind4}) as
\begin{equation}
    \begin{cases}
        \min_{\mathbf{r}(\cdot), \boldsymbol{\tau}(\cdot)} E_{\mathbf{h}} \left[
        \sum_{k=1}^K
        \tau_k(\mathbf{h})f_k(r_k(\mathbf{h})) \right]
        \\
        \mbox{subject to} \quad E_{\mathbf{h}} \left[ \tau_k(\mathbf{h})
        r_k(\mathbf{h})\right] = \bar{R}_k, \;\; k=1,\ldots, K, \\
        \hspace{2cm}  \sum_{k=1}^K \tau_k(\mathbf{h}) =1, \;\; \forall
        \mathbf{h}.
    \end{cases}
\end{equation}
Due to the feasibility and convexity of the power region
$\tilde{{\cal P}}(\mathbf{\bar{r}})$, the original problem
(\ref{E:ind1}) has a solution. Therefore, given any
$\boldsymbol{\mu}:=[\mu_1, \ldots, \mu_K]^T > \mathbf{0}$, there
exists a $\boldsymbol{\lambda}^*:= [\lambda_1^*, \ldots,
\lambda_K^*]^T > \mathbf{0}$, as well as optimal rate and time
allocation policies $\mathbf{r}(\cdot)$ and
$\boldsymbol{\tau}(\cdot)$ for (\ref{E:ind4}). It is clear that we
can again decompose (\ref{E:ind4}) into two sub-problems. Given
$\boldsymbol{\lambda}^*$, we first calculate $r_k^*(\mathbf{h})$
and $\tau_k^*(\mathbf{h})$ by solving (\ref{E:ind3}) and then we
determine the water-filling level $\boldsymbol{\lambda}^*$ by
satisfying individual rate constraints.

Since $f_k(r_k(\mathbf{h}))$ is convex in $r_k(\mathbf{h})$, it is
easy to show that it attains its minimum at $r_{k,\min}
(\mathbf{h}) = \left[ \log \lambda_k^* - \log \frac{\ln
2\mu_k}{h_k} \right]_+$. From the KKT conditions~\cite[Chapter
5]{convex}, it follows that for $\tau_k^*(\mathbf{h}) > 0$, we
should have $r_k^*(\mathbf{h})=r_{k,\min} (\mathbf{h})$. Next, we
show that a time allocation strategy $\boldsymbol{\tau} (\cdot)
\neq \boldsymbol{\tau}^*(\cdot)$, excluding the arbitrary time
sharing when functions $f_k(r_{k,\min}(\mathbf{h}))$ have multiple
minima, is strictly suboptimal.

Let us first consider the two-user case.
\begin{enumerate}
\item Suppose that for a fading state $\mathbf{h}$, we have
$f_1(r_{1,\min}(\mathbf{h}))<f_2(r_{2,\min}(\mathbf{h}))$ and
$\boldsymbol{\tau}(\mathbf{h})$ is a time allocation policy
different from $\boldsymbol{\tau}^*(\mathbf{h})$; i.e.,
$\tau_1(\mathbf{h})=1-\alpha$ and $\tau_2(\mathbf{h})=\alpha$ with
$\alpha > 0$. Consider the power cost
$J(\mathbf{h}):=(1-\alpha)f_1(r_{1,\min}(\mathbf{h}))+\alpha
f_2(r_{2,\min}(\mathbf{h}))$, where $r_{2,\min}(\mathbf{h}) \geq
0$ by definition.

\begin{enumerate}
\item If $r_{2,\min}(\mathbf{h})=0$, then we can let
$\tau_1'(\mathbf{h})=1$,
$r_1'(\mathbf{h})=r_{1,\min}(\mathbf{h})$, and
$\tau_2'(\mathbf{h})=r_2'(\mathbf{h})=0$ such that $J'(\mathbf{h})
= f_1(r_{1,\min}(\mathbf{h})) < J(\mathbf{h})$. It is clear that
$\tau_1'(\mathbf{h})r_1'(\mathbf{h}) > \tau_1
(\mathbf{h})r_{1,\min}(\mathbf{h})$, and since
$r_{2,\min}(\mathbf{h})=0$, $\tau_2'(\mathbf{h})r_2'(\mathbf{h}) =
\tau_2 (\mathbf{h})r_{2,\min}(\mathbf{h})=0$. Therefore, if
instead of $\boldsymbol{\tau}(\mathbf{h})$ we adopt
$\boldsymbol{\tau}'(\mathbf{h})$, we incur lower power cost
without violating the average individual rate constraints.

\item If $r_{2,\min}(\mathbf{h})>0$, we should have
$f_2^{(1)}(r_{2,\min}(\mathbf{h}))=0$. Let us consider the power
cost $J'(x,
\mathbf{h}):=(1-\alpha+x)f_1(r_{1,\min}(\mathbf{h}))+(\alpha-x)
f_2\left(\frac{\alpha}{\alpha-x}r_{2,\min}(\mathbf{h})\right)$,
and define $g(x):=J(\mathbf{h})-J'(x,\mathbf{h})$. Since $g(0)=0$
and $g^{(1)}(0)=
f_2(r_{2,\min}(\mathbf{h}))-f_1(r_{1,\min}(\mathbf{h}))
> 0$, there exists $\Delta \alpha \in (0, \alpha)$ such that
$g(\Delta \alpha)
> 0$. Therefore, we have $\tau_1'(\mathbf{h})=1-\alpha +\Delta
\alpha$, $r_1'(\mathbf{h})=r_{1,\min}(\mathbf{h})$,
$\tau_2'(\mathbf{h})=\alpha- \Delta \alpha$ and
$r_2'(\mathbf{h})=\frac{\alpha}{\alpha-\Delta
\alpha}r_{2,\min}(\mathbf{h})$, such that $J'(\Delta \alpha,
\mathbf{h}) < J(\mathbf{h})$. Since
$\tau_1'(\mathbf{h})r_1'(\mathbf{h})
> \tau_1 (\mathbf{h})r_{1,\min}(\mathbf{h})$ and $\tau_2'(\mathbf{h})r_2'(\mathbf{h}) = \tau_2
(\mathbf{h})r_{2,\min}(\mathbf{h})$, with
$\boldsymbol{\tau}'(\mathbf{h})$ and $\mathbf{r}'(\mathbf{h})$ we
can afford a smaller power without violating the average
individual rate constraints.
\end{enumerate}

\item For a fading state $\mathbf{h}$ satisfying
$f_1(r_{1,\min}(\mathbf{h}))>f_2(r_{2,\min}(\mathbf{h}))$ and
$\tau(\mathbf{h}) \neq \tau^*(\mathbf{h})$, it can be similarly
shown that we can have $J'(\mathbf{h}) < J(\mathbf{h})$ with
alternative resource allocation policies.
\end{enumerate}

Previous considerations show that the optimal resource allocation
policies should follow Theorem 4 for the two-user case. Similar
arguments extend readily to the general $K$-user case as well.

\subsection{Proof of Theorem 5}

\begin{enumerate}
\item[i)] From the optimal rate the time allocation policies, we
can directly verify the following fact. For all $k$, if the $k$th
component of $\boldsymbol{\lambda}$ increases while other
components remain fixed, $\bar{R}_k(\boldsymbol{\lambda})$
increases whereas $\bar{R}_i(\boldsymbol{\lambda})$ decreases for
$i \neq k$. More generally, for any subset ${\cal K}$, if we
increase $\lambda_k$ for all $k \in {\cal K}$ and hold the
remaining $\lambda_i$ fixed, $\bar{R}_i(\boldsymbol{\lambda})$
decreases for $i \in {\cal K}^C$, where superscript $C$ here
denotes set-complement.

\item[ii)] It can be easily verified that when
$\boldsymbol{\lambda}=\mathbf{0}$,
$\mathbf{\bar{r}}(\boldsymbol{\lambda})=\mathbf{0}$, and as
$\boldsymbol{\lambda} \uparrow \boldsymbol{\infty}$,
$\mathbf{\bar{r}}(\boldsymbol{\lambda}) \uparrow
\boldsymbol{\infty}$. This in turn implies that
\begin{enumerate}
\item $\forall \boldsymbol{\lambda}(0) > \mathbf{0}$, there exists
$\boldsymbol{\alpha} \leq \boldsymbol{\lambda}(0)$ for which
$\mathbf{\bar{r}}(\boldsymbol{\alpha}) \leq \mathbf{\bar{r}}$;

\item $\forall \boldsymbol{\lambda}(0) > \mathbf{0}$, there exists
$\boldsymbol{\beta} \geq \boldsymbol{\lambda}(0)$ for which
$\mathbf{\bar{r}}(\boldsymbol{\beta}) \geq \mathbf{\bar{r}}$.
\end{enumerate}

\item[iii)] Upon defining the mapping $\Lambda: \;
\boldsymbol{\lambda}(l) \rightarrowtail
\boldsymbol{\lambda}(l+1)$, we can readily verify from Lemma 3 and
i) that:
\begin{enumerate}
\item The vector $\boldsymbol{\lambda}^*$ is the unique fixed
point of the mapping $T$.

\item The mapping $\Lambda$ is order preserving; i.e.,
$\boldsymbol{\lambda} \leq \boldsymbol{\lambda}'$ $\Rightarrow$
$\Lambda(\boldsymbol{\lambda}) \leq
\Lambda(\boldsymbol{\lambda}')$.
\end{enumerate}

\item[iv)] We now establish one more fact.
\begin{enumerate}
\item If $\boldsymbol{\lambda}(0) \geq
\Lambda(\boldsymbol{\lambda}(0))$ and we define
$\boldsymbol{\lambda}(l):=\Lambda^l(\boldsymbol{\lambda}(0))$,
$l=0,1,\ldots$, then $\boldsymbol{\lambda}(l)$ is a decreasing
sequence and $\boldsymbol{\lambda}(l) \downarrow
\boldsymbol{\lambda}^*$.

\item If $\boldsymbol{\lambda}(0) \leq
\Lambda(\boldsymbol{\lambda}(0))$, then $\boldsymbol{\lambda}(l)$
is an increasing sequence and $\boldsymbol{\lambda}(l) \uparrow
\boldsymbol{\lambda}^*$.
\end{enumerate}

\emph{Proof:} We verify a) and b) as follows.
\begin{enumerate}
\item By preserving the order, we know that
$\boldsymbol{\lambda}(l)$ is decreasing. From ii), there exists a
$\boldsymbol{\alpha} \leq \boldsymbol{\lambda}(0)$ for which
$\mathbf{\bar{r}}(\boldsymbol{\alpha}) \leq \mathbf{\bar{r}}$. In
addition, by preserving the order, $\forall l$, we have
$\boldsymbol{\lambda}(l) \geq \Lambda^l(\boldsymbol{\alpha})$. But
since $\mathbf{\bar{r}}(\boldsymbol{\alpha}) \leq
\mathbf{\bar{r}}$, from i), we know that
$\Lambda^l(\boldsymbol{\alpha})$ is an increasing sequence. Hence,
$\{\boldsymbol{\lambda}(l)\}_{l=1}^{\infty}$ is decreasing and
bounded; thus, it must converge to the unique fixed point
$\boldsymbol{\lambda}^*$.

\item For $\boldsymbol{\lambda}(0) \leq
\Lambda(\boldsymbol{\lambda}(0))$, we can similarly prove the
claim.
\end{enumerate}
\end{enumerate}

Relying on i)-iv), we are ready to prove the theorem. Notice that
ii) guarantees the existence of points $\boldsymbol{\alpha}(0)$
and $\boldsymbol{\beta}(0)$ such that
\begin{equation}
\mbox{I) } \boldsymbol{\alpha}(0) \leq \boldsymbol{\lambda}(0)
\leq \boldsymbol{\beta}(0); \qquad \mbox{II) }
\mathbf{\bar{r}}(\boldsymbol{\alpha}(0)) \leq \mathbf{\bar{r}};
\qquad \mbox{III) } \mathbf{\bar{r}}(\boldsymbol{\beta}(0)) \geq
\mathbf{\bar{r}}.
\end{equation}

Defining
$\boldsymbol{\alpha}(l):=\Lambda^l(\boldsymbol{\alpha}(0))$ and
$\boldsymbol{\beta}(l):=\Lambda^l(\boldsymbol{\beta}(0))$, we know
from iv) that $\boldsymbol{\alpha}(l) \uparrow
\boldsymbol{\lambda}^*$ and $\boldsymbol{\beta}(l) \downarrow
\boldsymbol{\lambda}^*$. By preserving the order
$\boldsymbol{\alpha}(l) \leq \boldsymbol{\lambda}(l) \leq
\boldsymbol{\beta}(l)$, we have $\boldsymbol{\lambda}(l)
\rightarrow \boldsymbol{\lambda}^*$.

\subsection{Proof of Theorem 6}

For all CSI realizations $\mathbf{h}$, let
\begin{eqnarray}
    \boldsymbol{\tilde{\tau}}^* & := & [\tilde{\tau}_{1,0}^*, \ldots, \tilde{\tau}_{1,M_1}^*,
    \ldots, \tilde{\tau}_{K,0}^*, \ldots, \tilde{\tau}_{K,M_K}^*]^T,
    \\
    \boldsymbol{\tilde{\tau}} & := & [\tilde{\tau}_{1,0}, \ldots, \tilde{\tau}_{1,M_1},
    \ldots, \tilde{\tau}_{K,0}, \ldots, \tilde{\tau}_{K,M_K}]^T,
\end{eqnarray}
and define $\tilde{\tau}_k := \sum_{l=1}^{M_k}
\;\tilde{\tau}_{k,l}$. If $\mathbf{\bar{r}}$ is feasible, we have
$\boldsymbol{\lambda}^*$ for which (\ref{E:vlambda2}) is
satisfied. Let us also define $\mathbf{{\cal H}}_J:=\{ \mathbf{h}:
\; \mbox{$\left\{\varphi_k(\mathbf{h})\right\}_{k=1}^K$ have $J$
minima} \}$, for $J \in [1, K]$, and users $\{i_j\}_{j=1}^J$
having the ``best'' channels. Then $\forall
\boldsymbol{\tilde{\tau}} \neq \boldsymbol{\tilde{\tau}}^*$, we
have
\begin{eqnarray}
    \nonumber
    \lefteqn{\sum_{k=1}^K \; \lambda_k^* \left(E_{\mathbf{h}} \left[
    \sum_{l=1}^{M_k} \tilde{\tau}_{k,l} \rho_{k,l} \right] - \bar{R}_k \right)} \\
    \nonumber
    & = &
    \sum_{k=1}^K \; \lambda_k^* \left(E_{\mathbf{h}} \left[
    \sum_{l=1}^{M_k} \tilde{\tau}_{k,l} \rho_{k,l} \right] - E_{\mathbf{h}} \left[
    \sum_{l=1}^{M_k} \tilde{\tau}_{k,l}^* \rho_{k,l} \right] \right) \\
    \nonumber
    & = & E_{\mathbf{h} \in \mathbf{{\cal H}}_1} \left[ \sum_{k \neq i, \; k=1}^K \;
        \left(\lambda_k^* \sum_{l=1}^{M_k} \tilde{\tau}_{k,l} \rho_{k,l} \right) +
        \lambda_i^* \left(\sum_{l=1}^{M_i} \tilde{\tau}_{i,l} \rho_{i,l}
        \; - \rho_{i,l_i^*} \right)  \right] \\
    & & + \sum_{J=2}^{K} \; E_{\mathbf{h} \in \mathbf{{\cal H}}_J} \left[ \sum_{k \neq i_j, \; k=1}^K \;
        \left(\lambda_k^* \sum_{l=1}^{M_k} \tilde{\tau}_{k,l} \rho_{k,l} \right) +
        \sum_{j=1}^J \; \lambda_{i_j}^*
        \left(\sum_{l=1}^{M_{i_j}} \tilde{\tau}_{i_j,l} \rho_{i_j,l}
        \; - \tau_j^* \rho_{i_j,l_{i_j}^*} \right) \right].
    \label{E:rho}
\end{eqnarray}

\begin{enumerate}
\item $\forall \mathbf{h} \in \mathbf{{\cal H}}_1$, suppose that
user $i$ is selected by $\boldsymbol{\tilde{\tau}}^*$ as the user
with the best channel. Since $\mu_k \gamma_{k,l_k^*}/h_k \leq
\lambda_k^* < \mu_k \gamma_{k,l_k^*+1}/h_k$, $k \in [1, K]$, we
have
$\frac{C_{k,l_k^*}-C_{k,l_k^*-1}}{\rho_{k,l_k^*}-\rho_{k,l_k^*-1}}
\leq \lambda_k^* <
\frac{C_{k,l_k^*+1}-C_{k,l_k^*}}{\rho_{k,l_k^*+1}-\rho_{k,l_k^*}}$.
Therefore,
\begin{equation}
    \rho_{k,l_k^*-1} \leq \rho_{k,l_k^*} + (C_{k,l_k^*-1}-C_{k,l_k^*})/\lambda_k^*, \quad \quad
    \rho_{k,l_k^*+1} < \rho_{k,l_j^*} + (C_{k,l_k^*+1}-C_{k,l_k^*})/\lambda_k^*.
\end{equation}
By the convexity of $\Upsilon_k (x)$, we have $\mu_k
\gamma_{k,l}/h_k < \lambda_k^* < \mu_k \gamma_{k,l'}/h_k$,
$\forall l \in [1, l_k^*-1]$, $\forall l' \in [l_k^*+1, M_k]$.
Hence, $\forall l \neq l_k^*$, $l \in [1, M_k]$, we have
$\rho_{k,l} < \rho_{k,l_k^*} + (C_{k,l}-C_{k,l_k^*})/\lambda_k^*$.
Then the first sum of the last equality in (\ref{E:rho}) will be
\begin{eqnarray}
    \nonumber
    \lefteqn{E_{\mathbf{h} \in \mathbf{{\cal H}}_1} \left[ \sum_{k \neq i, \; k=1}^K \;
        \left(\lambda_k^* \sum_{l=1}^{M_k} \tilde{\tau}_{k,l} \rho_{k,l} \right) +
        \lambda_i^* \left(\sum_{l=1}^{M_i} \tilde{\tau}_{i,l} \rho_{i,l}
        \; - \rho_{i,l_i^*} \right)  \right]} \\
    \nonumber
    & \leq & E_{\mathbf{h} \in \mathbf{{\cal H}}_1} \left[ \sum_{k \neq i, \; k=1}^K \;
        \left( \sum_{l \neq l_k^*, \; l=1}^{M_k} \tilde{\tau}_{k,l} (C_{k,l}-C_{k,l_k^*}) \;
        + \lambda_k^* \tilde{\tau}_k \rho_{k,l_k^*} \right) \right. \\
    \nonumber
        & & \hspace{1.5cm} \left. +
        \sum_{l \neq l_i^*, \; l=1}^{M_i} \tilde{\tau}_{i,l} (C_{i,l}-C_{i,l_i^*})
        + \lambda_i^* \tilde{\tau}_i \rho_{i,l_i^*} - \lambda_i^* \rho_{i,l_i^*}
        \right] \\
    \nonumber
    & = & E_{\mathbf{h} \in \mathbf{{\cal H}}_1} \left[ \sum_{k \neq i, \; k=1}^K \;
        \left( \sum_{l=1}^{M_k} \tilde{\tau}_{k,l} C_{k,l} \; - \tilde{\tau}_k (C_{k,l_k^*}
        - \lambda_k^* \rho_{k,l_k^*}) \right) \right. \\
    \nonumber
        & & \hspace{1.5cm} \left. +
        \sum_{l=1}^{M_i} \tilde{\tau}_{i,l} C_{i,l} \; - \tilde{\tau}_i (C_{i,l_i^*}
        - \lambda_i^* \rho_{i,l_i^*}) - \lambda_i^* \rho_{i,l_i^*}
        \right] \\
    & \leq & E_{\mathbf{h} \in \mathbf{{\cal H}}_1} \left[ \sum_{k=1}^K \;
        \sum_{l=1}^{M_k} \tilde{\tau}_{k,l} C_{k,l} \; - \sum_{k=1}^K \tilde{\tau}_k (C_{i,l_i^*}
        - \lambda_i^* \rho_{i,l_i^*})  - \lambda_i^* \rho_{i,l_i^*}
        \right] \label{E:amcpoly} \\
    & = & E_{\mathbf{h} \in \mathbf{{\cal H}}_1} \left[ \sum_{k=1}^K \;
        \sum_{l=1}^{M_k} \tilde{\tau}_{k,l} C_{k,l} \; - C_{i,l_i^*}
        \right];
    \label{E:HA1}
\end{eqnarray}
where we used the allocation policies $\forall k \neq i$,
$C_{k,l_k^*}- \lambda_k^* \rho_{k,l_k^*} > C_{i,l_i^*}-
\lambda_i^* \rho_{i,l_i^*}$ to obtain inequality
(\ref{E:amcpoly}).

\item $\forall \mathbf{h} \in \mathbf{{\cal H}}_J$, $J>1$, suppose
that users $i_j$ are selected by $\boldsymbol{\tilde{\tau}}^*$ as
the users with best channels, and define
$C_{\min}:=C_{i_j,l_{i_j}^*}- \lambda_{i_j}^*
\rho_{i_j,l_{i_j}^*}$, $j=1,\ldots, J$. Noticing that $\forall k
\neq i_j$, $C_{k,l_k^*}- \lambda_k^* \rho_{k,l_k^*} > C_{\min}$
and that $\forall l \neq l_k^*$, $l \in [1, M_k]$, $\rho_{k,l}
\leq \rho_{k,l_k^*} + (C_{k,l}-C_{k,l_k^*})/\lambda_k^*$, we have
\begin{eqnarray}
    \nonumber
    \lefteqn{E_{\mathbf{h} \in \mathbf{{\cal H}}_J} \left[ \sum_{k \neq i_j, \; k=1}^K \;
        \left(\lambda_k^* \sum_{l=1}^{M_k} \tilde{\tau}_{k,l} \rho_{k,l} \right) +
        \sum_{j=1}^J \; \lambda_{i_j}^*
        \left(\sum_{l=1}^{M_{i_j}} \tilde{\tau}_{i_j,l} \rho_{i_j,l}
        \; - \tau_j^* \rho_{i_j,l_{i_j}^*} \right) \right]} \\
    \nonumber
    & \leq & E_{\mathbf{h} \in \mathbf{{\cal H}}_J} \left[ \sum_{k \neq i_j, \; k=1}^K \;
        \left( \sum_{l=1}^{M_k} \tilde{\tau}_{k,l} C_{k,l} \; - \tilde{\tau}_k (C_{k,l_k^*}
        - \lambda_k^* \rho_{k,l_k^*}) \right) \right. \\
    \nonumber
        & & \hspace{1.5cm} \left. +
        \sum_{j=1}^J  \left(\sum_{l=1}^{M_{i_j}} \tilde{\tau}_{i_j,l} C_{i_j,l} \; - \tilde{\tau}_{i_j} (C_{i_j,l_{i_j}^*}
        - \lambda_{i_j}^* \rho_{i_j,l_{i_j}^*}) - \tau_j^* \lambda_{i_j}^* \rho_{i_j,l_{i_j}^*} \right)
        \right] 
\end{eqnarray}
\begin{eqnarray}
    \nonumber
    & \leq & E_{\mathbf{h} \in \mathbf{{\cal H}}_J} \left[ \sum_{k=1}^K \;
        \sum_{l=1}^{M_k} \tilde{\tau}_{k,l} C_{k,l} \; - C_{\min}  -
        \sum_{j=1}^{J} \tau_j^* \lambda_{i_j}^* \rho_{i_j,l_{i_j}^*}
        \right] \\
    \nonumber
    & = & E_{\mathbf{h} \in \mathbf{{\cal H}}_J} \left[ \sum_{k=1}^K \;
        \sum_{l=1}^{M_k} \tilde{\tau}_{k,l} C_{k,l} \; - \sum_{j=1}^{J} \tau_j^* (C_{i_j,l_{i_j}^*}
        - \lambda_{i_j}^* \rho_{i_j,l_{i_j}^*}) -
        \sum_{j=1}^{J} \tau_j^* \lambda_{i_j}^* \rho_{i_j,l_{i_j}^*}
        \right] \\
    & = & E_{\mathbf{h} \in \mathbf{{\cal H}}_J} \left[ \sum_{k=1}^K \;
        \sum_{l=1}^{M_k} \tilde{\tau}_{k,l} C_{k,l} \; - \sum_{j=1}^{J} \tau_j^* C_{i_j,l_{i_j}^*}
        \right].
    \label{E:HAK}
\end{eqnarray}
\end{enumerate}

Substituting (\ref{E:HA1}) and (\ref{E:HAK}) into (\ref{E:rho}),
we have
\begin{eqnarray}
    \nonumber
    \lefteqn{\sum_{k=1}^K \; \lambda_k^* \left(E_{\mathbf{h}} \left[
    \sum_{l=1}^{M_k} \tilde{\tau}_{k,l} \rho_{k,l} \right] - \bar{R}_k \right)}\\
    \nonumber
    & \leq &
    \sum_{J=1}^{K} E_{\mathbf{h} \in \mathbf{{\cal H}}_J} \left[ \sum_{k=1}^K \;
        \sum_{l=1}^{M_k} \tilde{\tau}_{k,l} C_{k,l} \; - \sum_{j=1}^{J} \tau_j^* C_{i_j,l_{i_j}^*}
        \right] \\
    & = & E_{\mathbf{h}} \left[ \sum_{k=1}^K \;
        \sum_{l=1}^{M_k} \tilde{\tau}_{k,l} C_{k,l} \right] -
        E_{\mathbf{h}} \left[ \sum_{k=1}^K \;
        \sum_{l=1}^{M_k} \tilde{\tau}_{k,l}^* C_{k,l} \right].
\end{eqnarray}

Therefore, $\forall \boldsymbol{\tilde{\tau}} \neq
\boldsymbol{\tilde{\tau}}^*$, if $\boldsymbol{\tilde{\tau}}$
satisfies individual rate constraints $E_{\mathbf{h}}
\left[\sum_{l=1}^{M_k} \tilde{\tau}_{k,l} \rho_{k,l} \right] \geq
\bar{R}_k$, we have
\begin{equation}
    E_{\mathbf{h}} \left[ \sum_{k=1}^K \;
   \sum_{l=1}^{M_k} \tilde{\tau}_{k,l} C_{k,l} \right]  \geq
    E_{\mathbf{h}} \left[ \sum_{k=1}^K \;
    \sum_{l=1}^{M_k} \tilde{\tau}_{k,l}^* C_{k,l} \right].
\end{equation}
Hence, $\boldsymbol{\tilde{\tau}}^*$ is the optimal solution to
(\ref{E:optamc3}) and consequently the corresponding
$\mathbf{r}^*$ and $\boldsymbol{\tau}^*$ are the optimal solutions
to (\ref{E:optamc2}).

Similar to Lemma 3, we can show that $\boldsymbol{\lambda}^*$ is
almost surely unique. Define ${\cal G}$ as the set of all feasible
rate vectors. If $\mathbf{\bar{r}}$ is feasible, there must be a
boundary point $\mathbf{\bar{r}}_g$ of ${\cal G}$ for which
$\mathbf{\bar{r}} \leq \mathbf{\bar{r}}_g$. Let
$\boldsymbol{\lambda}_g$ denote the Lagrange multiplier
corresponding to $\mathbf{\bar{r}}_g$. Then for
$\boldsymbol{\lambda}(0) \leq \boldsymbol{\lambda}_g$, there
exists $\boldsymbol{\beta} \geq \boldsymbol{\lambda}(0)$ for which
$\mathbf{\bar{r}}(\boldsymbol{\beta}) \geq \mathbf{\bar{r}}$. With
this guarantee replacing the counterpart ii)-b) and following the
lines in the proof of Theorem 5 (Appendix F), we can show that
$\boldsymbol{\lambda}^*$ can be iteratively computed by Algorithm
4 for any positive initialization $\boldsymbol{\lambda}(0) \leq
\boldsymbol{\lambda}_g$.

\newpage
\bibliographystyle{amsplain}



\vspace{0.5cm}

\newpage

\end{document}